\documentclass[10pt,doublecolumn,journal]{IEEEtran}
\usepackage{amsmath,amssymb,amsfonts}
\usepackage{tabu}
\usepackage{amsthm} 
\usepackage{amsmath, nccmath}
\usepackage{float}
\usepackage{mathtools}
\usepackage{graphicx}
\usepackage{graphics}
\usepackage{caption}
\usepackage{subcaption}
\usepackage{color}
\usepackage{comment}
\usepackage{setspace} 
\usepackage{caption}
\usepackage{subcaption}
\usepackage{hyperref}
\usepackage{footnote}
\usepackage{subcaption}

\begin{document}

\title{A Vectored Fragmentation Metric for Elastic Optical Networks}
\author{Anjali Sharma, Varsha Lohani and Yatindra Nath Singh \\  \textit{Department of Electrical Engineering, \\ Indian Institute of Technology Kanpur} }
\date{2019}

\maketitle

\begin{abstract}
When circuits are set up and dismantled dynamically in elastic optical networks, spectrum tends to become fragmented in the fiber links. The fragmentation limits the available path choices and may lead to significant blocking of connection requests. There are two types of fragmentation in the network spectrum- in the links due to contiguity constraints and over the paths due to continuity. Study of fragmentation and its management is essential to operate the networks efficiently. This paper proposes a \textit{vectored fragmentation metric} for characterizing the fragmentation, which covers both types of fragmentation. We discuss the characteristics of this metric in both transient and steady-state of the dynamic network. We also test the proposed metric for connection requests' granularity range, arrival rates and holding times, to establish functionality of this metric. We also compare the link-based fragmentation metric with our Vectored Fragmentation Metric to understand the better representation.   
\end{abstract}


\section{INTRODUCTION}


Over last few decades, optical fiber communications have emerged as the core of communication infrastructure across the globe. The reason for this is capability of long distance transmission using optical fiber due to its inherently large bandwidth distance product. Further, availability of huge bandwidth has fuelled innovation and development of applications which use this bandwidth and consequently further fuel the bandwidth demand. According to the Cisco Annual Internet report (2018-2023), nearly two-thirds of the global population will have access to the internet and mobile connectivity by 2023 \cite{I1}.

This growing demand has led to intense research in wavelength routed networks, optical circuit, packet and burst switching. Elastic optical networks (EONs) have also been posed as a possible way of moving ahead from WDM (Wavelength Division Multiplexed) networks leading to more efficient and hence possibly more economical networks.


WDM networks allocate fixed bandwidth channels to any demand received.  Typically 25 GHz or 50 GHz channels are used in WDM networks. In a network with 50 GHz channels, if a bandwidth request of 5 GHz\footnote{in terms of spectral grid requirement} arrives at an ingress node, the network has no option except to allocate a 50 GHz channel. Elastic optical networks are conceived to resolve this limitation. In EONs, the lowest bandwidth that can be fulfilled by a single channel with 100 percent efficiency is 12.5 GHz \cite{I2}. Thus, if received by an Elastic Optical Network, the same 5 GHz request will get only 12.5 GHz allocated and not 50 GHz. Finer bandwidth allocation granularity allows more efficient use of resources. Thus we expect better and efficient use of the existing optical networks and thus increased effective capacity catering to more bandwidth demand. The tutorial paper \cite{I3} and \cite{I4} cover the enabling technologies for EON at hardware (transceivers and switches) and network level (resource allocation schemes).  


Instead of defining a fixed bandwidth channel of 50 GHz as in WDM, EON defines a bandwidth slice\footnote{a group of slices is considered as a bandwidth slot in this paper.} as a basic unit. In order to cater to large bandwidth, multiple adjacent slices in the spectrum can be allocated together. Thus, bandwidth demands can be accommodated in a given spectrum by allocating bandwidth in integral multiples of the basic slice unit, or Frequency Slot Unit \cite{I5}. As a consequence, the wastage of spectrum is reduced drastically as compared to WDM networks. However, this strategy of elastically allocating several adjacent slices depending on demand also leads to another problem, i.e., fragmentation. Sometimes bandwidth, despite being available, cannot be allocated as it is not contiguous.

This paper is investigating elastic optical networks in reference to improving the methods for maintaining the efficiency, and hence possibly improving the economics of using them. The organization of the paper is as follows. In Section II, we discuss the basic concept of resource provisioning and issues related to it in EONs.  Section III discusses the concept of the vectored fragmentation metric. In the subsections, we discuss the metric's formulation along with an example of fragmentation level calculation. We also evaluate the properties of the metric and its application. Section IV presents the assessment of the proposed fragmentation metric using simulation results. Section V closes the paper with a general conclusion, and future challenges in the study of vectored fragmentation metric.

\section{Basic Concepts in Elastic Optical Networks}
\subsection{ Routing and Spectrum Assignment}

\begin{figure}[!t]
        \includegraphics[width= \linewidth]{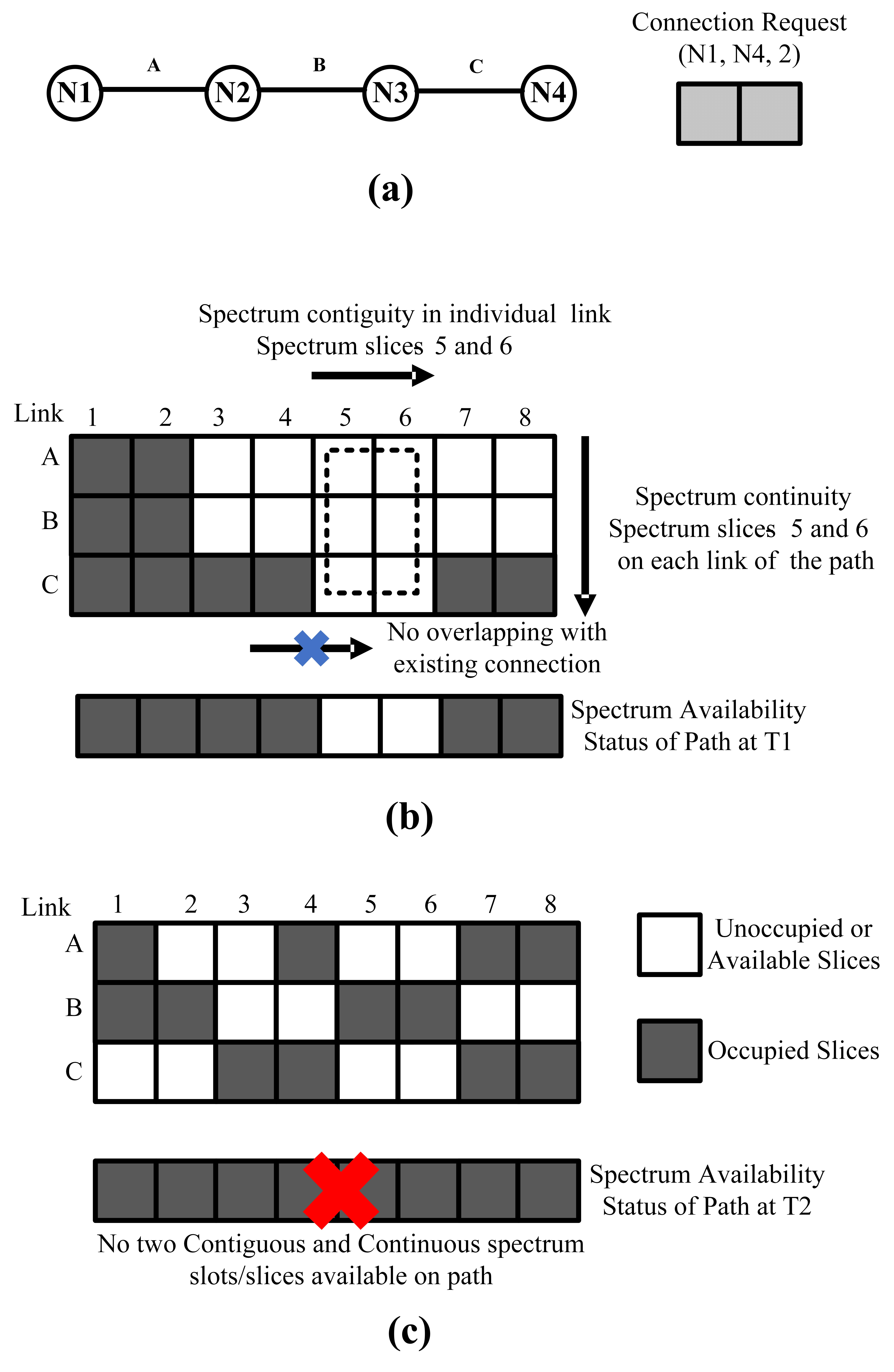}
        \caption{RSA Constraints example. (a) A connection request between N1 and N4 with two spectrum slices requirement, (b) given a network spectrum scenario at T1, and the slots on a path complying with the constraints, (c) Non- availability of slices for the same connection request at T2.}
        \label{fig:fig1}
\end{figure}

In WDM networks, the connection requests on their arrival, are set up through the nodes and the links using a wavelength channel. Routing and Wavelength Assignment problem ensures that the connection request is set up on a selected path, and its bandwidth requirement from the source (also called ingress) to the destination (also called egress) is satisfied. The allocated wavelength channel may be selected using First-fit or Last-fit assignment rules. In case conversion of wavelength is not feasible in the path, then the path has to be on the same wavelength. If it cannot be done, path setup is declared infeasible. Commonly, we should set up the connection request on a shortest path using the shortest path algorithm. If we do not enforce the shortest path constraint, it will still be possible to set up the connection request on the same wavelength (when no converters are there) through a longer route.

Routing and wavelength assignment problem need to be solved in a reasonable time to set up paths if feasible. For an operating network, the problem becomes an incremental optimization problem. When the requests arrive, the best possible path is set up while trying to reduce the blocking probability.

The same problem gets extended to Elastic Optical Network Scenario with another constraint of contiguity in addition to continuity \cite{c2}. Spectral continuity constraint sets the rule that the same part of the spectrum is allocated all through the path unless there are waveband shifters at some intermediate nodes where there is an option to shift the allocated spectral band. Spectral contiguity constraint dictates that the slices allocated to the path are together forming a continuous band in the network spectrum. Due to these constraints, the routing and wavelength assignment problem now becomes an RSA (routing and spectral assignment) problem. An example of how a connection request is provisioned under the RSA constraints in a linear network is shown in fig.~\ref{fig:fig1}(a)-(b). If the slot satisfying the constraints are available, then the connection request is provisioned and set up fig.~\ref{fig:fig1}(b). If sufficient resources are not available or not complying the RSA constraints, for a connection request, then that connection request is blocked fig.~\ref{fig:fig1}(c). In this paper, we first discuss this problem and the limitations associated with it.

An EON can deploy RSA strategies for both static traffic as well as dynamic traffic scenarios. In static traffic, paths of all the connection requests are known a priori. Hence, the lightpath allocation and configuration of switches are computed and configured beforehand. In a dynamic traffic scenario, connection requests arrive as well as depart randomly. Appropriate stochastic models can characterize the arrival and departure. The RSA decides the allocation of the spectrum slices and path after the arrival of connection. At this step, there may arise a situation where the resources are not available. In that case, connection request is rejected (blocked). In a real-life situation, most connection requests arrive dynamically; thus, we characterize the network performance in terms of blocking probability for given arrival rates and connection duration statistics.

Ideally, in both static or dynamic traffic scenarios, all the paths that satisfy the constraints are identified, and then the best path (which can be a shortest path) is chosen. We are assuming that connections are arriving one at a time sequentially. If path requests contend for the same link, the request arriving earlier gets precedence, and the second one is either setup through an alternate path or gets blocked. Next, we discuss spectrum fragmentation issue and related inefficiencies in EON.

\subsection{Spectrum Fragmentation}

In dynamic traffic scenarios, the connection requests with different bandwidth requirements keep coming and leaving. As the RSA constraints of contiguity and continuity are to be satisfied while setting up a connection, the situations arise when the required number of slices are available, but still, the path cannot be set up. The path is not available there because either the continuity or the contiguity constraint is not satisfied. Though if all the existing connections could have been reorganized, then some of the refused connections could have been set up. Such situations depict the network's fragmented state, and the reorganization is the desirable defragmentation process in the network.

Fragmentation can be understood to be happening due to a lack of continuity or contiguity fulfillment of spectrum slices. The fragmentation brings spectrum inefficiency and degrades network performance. To address this fragmentation issue, we identify two types of fragmentation- the one due to the non-continuity of available resources on a path and the second one due to the non-contiguity of available resources on a link. Consider the scenario as shown in fig.~\ref{fig:fig1}(c). Here a connection request for two slices from node N1 to node N4 cannot be setup. In each of the individual links, i.e., A, B, and C, more than two slices are available, and still, due to continuity constraint, the request cannot be satisfied. The connections with shorter hops and lesser slice requests will have higher chances of getting through, i.e., will have lesser blocking probability. 

\begin{figure}[!t]
   \begin{center}
       \includegraphics[width=\linewidth]{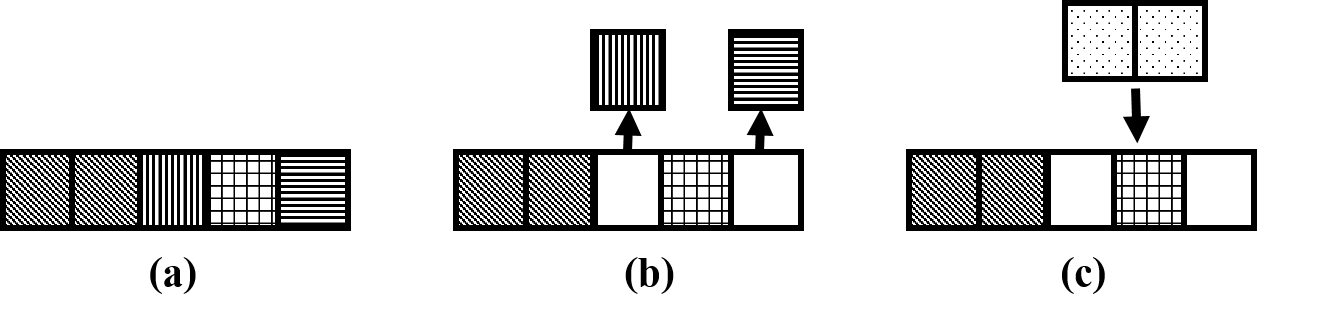}
        \caption{(a) Connection requests occupying spectrum resources at a time $t_1$, then (b) two connection requests depart at $t_2$, two spectrum slices become available. (c) A new connection request with 2 spectrum slice requirement arrives at $t_3$, but cannot be accommodated as available resources are not contiguous.}
        \label{fig:fig2}
    \end{center}
\end{figure}

Fragmentation also leads to unfair treatment to connections requesting higher bandwidth, as in a fragmented situation, the low bandwidth connection has a higher probability of getting through (fig.~\ref{fig:fig2}). A few works with Markov Chain (MC) model tried to characterize the fragmentation in network by using a single channel and a super-channel services, \cite{MC1},\cite{MC2}. The blocking probability plot of individual services shows how the single channel requests rob off the resources from super-channel requests. The fragmentation management approaches are of utmost importance to deal with the inefficiencies and unfairness in the system.



There are two approaches to manage the fragmentation: one with no reorganizations (non-defragmentation) and the other with periodic reorganization (i.e., a defragmentation procedure). In the first approach, i.e. non-defragmentation, the objective is to operate the network to minimize fragmentation. In the periodic defragmentation approach, whenever the fragmentation increases, reorganization of existing connections is done to allow for setting up of new connections.

The non-defragmentation approach cannot guarantee the $100\%$ fragmentation less operation of the network. While the defragmentation approach requires some interruption in the existing connections when their reorganization occurs. But the defragmentation approach is expected to give better utilization of resources.


Both the management strategies can reduce the consumption network resources. But for the second strategy, it is important to defragment the network at right instants. We need to define a methodology to quantify the fragmentation level and use it to identify the time instants to defragment the network. The methodology should also help decipher the causes behind the fragmentation.


There have been several attempts to quantify the
fragmentation. These are either based on fragmentation in links' spectrum, or fragmentation through paths, or throughout the network.

Some of the link-based fragmentation metrics which exploit the spectrum status in links have been discussed in \cite{c3, FMs, FM1}. Some of the link-based fragmentation measures are - the external fragmentation \cite{c3}, the Shannon entropy-based measure \cite{c3}, the Access Blocking probability-based measure \cite{c3}, the utilization entropy \cite{FMs}, the high-slot mark \cite{FMs} and the spectrum compactness \cite{FM1}. The metrics discussed in the cited works give an abstract treatment to the fragmentation measure. They tend to ignore some of the crucial aspects of spectrum status, e.g., small fragments, and cannot differentiate between different spectrum scenarios.

Some of the proposed fragmentation measures are dependent on connection requests and paths requested by them. Pederzolli {\em et al.} \cite{c4} have proposed a path-based fragmentation metric. It accounts for both wasted and unusable slots (equivalent to slices in this paper) in a path, capturing the onset of fragmentation for a specific connection request. The authors use this metric in the RSA, to find path-slots combination for a connection request. The selected path-slots combination should result in lower overall fragmentation value. So, appropriate path selection is precursor to fragmentation, if any.  While this metric captures fragmentation due to both continuity and contiguity both, but it is path/connection request specific. 
In \cite{c41}, authors used two network-level metrics, one associated with the contiguity aspect and the other associated with network utilization. The first metric uses the link consecutiveness aspect in all links at specific observation periods (time-weighted) to decide on a network's fragmentation level. The second metric is also a time-weighted network utilization metric. It accounts for the unavailability of links due to high load, which should not be a reason for fragmentation. The authors in this work also gave a perception of unfairness i.e., high bandwidth requests are more likely to get blocked due to fragmentation. 


There have also been attempts to devise the fragmentation-aware routing and spectrum allocation techniques \cite{c5, c6}. The metric can provide information about the onset of fragmentation shortly after the fragmentation has set in. In both scenarios, one can update routing or spectrum assignment strategy one the onset of fragmentation is detected. For the fragmentation awareness part, various spectrum parameters in the links of the network, are monitored. Though, these parameters may not have a direct link to the fragmentation. In the given schemes, different spectrum allocation strategies are updated to achieve maximum request acceptance even in the presence of fragmentation. The authors in \cite{c11} made another interesting attempt where they accommodate the connection requests only if they do not lead to fragmentation in the future. However, it lead to selective acceptance and, in turn, unfairness in the system. The common aspect in all of the above fragmentation management approaches is the use of fragmentation metric to either routing (deadlock-avoidance) or spectrum allocation (to achieve lowest contiguity ratio). The contiguity ratio is the ratio of maximum contiguously available slices to total available slices in a link.

This paper presents a two-dimensional fragmentation metric, which gives an absolute value for the fragmentation status for the whole network.  We call it a vectored fragmentation metric (VFM). It has two fragmentation components, one covering the fragmentation due to continuity constraint over a number of links, and the other, covering the fragmentation due to contiguity constraint within the links. These are calculated independently of each other.

\section{A Vectored Fragmentation Metric}

A fragmentation metric quantifies the level of fragmentation in the spectrum. The fragmentation level can be high or low, depending on the vacant bandwidth slice positioning in the spectrum on the links and its continuity over the links forming a continuous path. The standard performance metrics for the network, e.g., blocking performance for new arriving connections, are expected to reduce for the lesser value of fragmentation metric for the same spectrum utilization.

\begin{figure}[!t]
\centering
\includegraphics[width=0.8\linewidth,keepaspectratio]{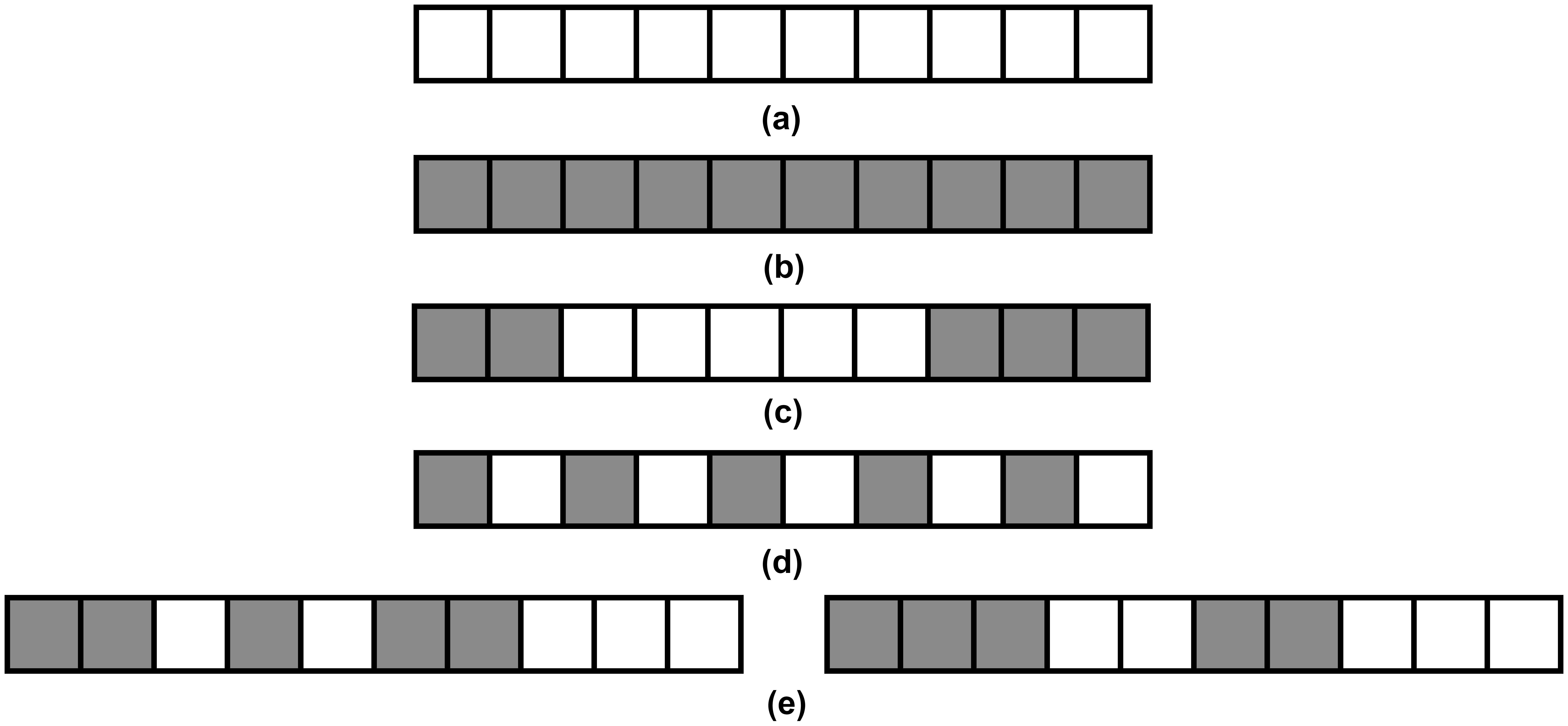}
\caption{Types of fragmentation cases.}
\label{fig:fig3}
\end{figure}

We would like to have a single metric that can be entrusted with representing the fragmentation level, considering both contiguity and continuity aspects. Several metrics have been proposed in the literature using different network spectrum characteristics (e.g., link spectrum status). However, they fail to quantify the fragmentation satisfactorily, e.g., the inability to identify smaller fragments. Based on earlier studies, we enumerate a few cases which a useful fragmentation metric should be able to identify (fig.~\ref{fig:fig3}).

\begin{enumerate}
    \item {Case (a)- Where all slices are free (No fragmentation)}
    \item {Case (b)- Where all slices are busy (Case of no fragmentation; blocking due to resource unavailability)}
    \item {Case (c)- Where free slices are contiguous (No fragmentation)}
    \item {Case (d)- Where free slices are lost/unusable (absolute fragmentation)}
    \item {Case (e)- Higher the fragmentation in the spectrum, larger is the metric value (relative fragmentation).}
\end{enumerate}

Blocking of a connection can happen due to unavailability of contiguous bandwidth slices in one or more links forming the path over which connection is to be setup. Thus, the fragmentation due to both continuity and contiguity are a major contributors to the blocking of connection requests. It will happen due to improperly managed RSA.

We formulate a vectored fragmentation metric which provides an absolute single value for the fragmentation level while considering both continuity as well as contiguity. The proposed metric takes into account the fragmentation in individual links (by considering largest number of contiguous slices) and fragmentation over multiple links forming a path, by finding the maximum number of continuously free links for each spectrum slice.   

Operationally, We can assume a centralized network controller (like SDN controller) which interacts with all the routers to gather all the status, and setup the paths for the arriving connection demands. It will have extensive status of components and can help in making better decisions \cite{c9}. The central controller can determine the vectored fragmentation metric using the individual links' status and continuity of available slices across the specified paths. Though the specified path will have impact on the computed metric.

VFM consists of $\alpha$- and $\beta$-components.

\begin{itemize}
    \item{ \textbf{$\alpha$-component}: which covers the fragmentation due to non-contiguity of available spectrum slices in individual links across the whole network.}
    \item{ \textbf{$\beta$-component}: which covers the fragmentation due to non-continuity of available spectrum slices over the longest paths in the network covering all the links.}
\end{itemize}

\subsection{Formulation of metric}

The $\nu$ is the fragmentation indicator also called Vector Fragmentation Metric (VFM). It is resultant of $\alpha$ and $\beta$  components. For the $\alpha$- component, the maximum contiguous slices in a link are taken up against total available slices in that link. Ideally, if all the available slices form a single contiguous slot, then there is no fragmentation. An important assumption is that at least one spectrum slice is available in the network spectrum. More than one spectrum slices scattered is the main cause of fragmentation. For the $\beta$-component, continuity of a single slice over a path is taken into account. If possible a single longest path (a cycle), or otherwise, multiple paths are used to check continuity of each slice index over all the links in each path. When single path is not feasible, minimum path(s) are chosen in such a way that no link is repeated and status of all the links are covered. Then, the continuity of all the spectrum slice indices (which are available at least in one of the links on the path) is used. The maximum number of continuous hops where slices (of a particular index) are available, is taken against the total available (unused) hops for that spectrum slice over the whole path. The calculation of the components and the fragmentation indicator is done as in equations \ref{1} - \ref{3}. 

\begin{equation} \label{1}
    \alpha = \frac{1}{|EL|}. \sum_{i=1}^{EL} \frac{CG_{i}}{SS_i}
\end{equation}

\begin{equation}\label{2}
    \beta = \frac{1}{|P|} . \sum_{i=1}^{P} \frac{1}{E_{i}} \sum_{j=1}^{E_{i}} \frac{CN_{j}^{i}}{AS_{j}^{i}} 
\end{equation}

\begin{equation}\label{3}
    VFM = \nu = \sqrt{\alpha^2 + \beta^2}
\end{equation}

where
\begin{itemize}
    \item[] $EL$ = set of links with at least one empty (or available) spectrum slice,
    \item[] $SS_i$ = total number of available spectrum slices in the $i^{th}$ link,
    \item[] $CG_{i}$ = Maximum number of contiguous spectrum slices available on $i^{th}$ link in the network spectrum, 
    \item[] $P$ = set of multiple paths in the network specified for the path continuity aspect, covering all the links
    \item[] $E_i$ = total spectrum slice indices in the network spectrum with at least one empty spectrum slice anywhere on the path in the $i^{th}$ path,
    \item[] $AS_j^i$= Number of hops  where slices on $j^{th}$ index is free in the $i^{th}$ path,
    \item[] $CN_{j}^{i}$ = Maximum number of continuous hops with available $j^{th}$ spectrum slices in $i^{th}$ multihop path, 
    \item[] $TSS$ = total number of spectrum slices in all the links (includes both available and occupied),
    \item[] $H_p$= Number of hops on $p^{th}$ path,
    \item[] $L$= set of links in the network.
\end{itemize}

This formulation calculates the fragmentation level in the network spectrum using the available spectrum slices. If there are no available spectrum slices, then no fragmentation exists.  We emphasize on longest path selection (or a Hamiltonian path) for $\beta-$component as it ensures that we get a network-wide beta-component and its value is not independent of path-specific continuity aspect. As the $\beta$- component checks continuity of all spectrum slices individually, i.e., the component decides on non-continuity fragmentation from a single slice index's point of view. Therefore, even if several single spectrum slices (at high spectrum utilization) are available on multiple indices along the path, which is a case of fragmentation, it gets ignored in $\beta$-component calculation. Later, we also intend to use different path combinations in both Hamiltonian and non-Hamiltonian network to see if the choice of paths has any impact on the defragmentation trigger and hence the blocking performance. Another important question is, if using a single longest or multiple paths with no or minimum repetition of links has any impact on the $\beta-$component's contribution in the fragmentation level? As per our understanding, we would like to surmise here that if one considers static routing, then using multiple paths could be a better for $\beta$ calculation. We hypothesize so because in static routing case the paths are fixed and hence easy to rely on. However, in dynamic routing scenario, the paths are not fixed and the controller looks for the best possible path available. So in place of using large multiple paths we try to use a a single longest path or multiple longest paths which can preserve the flow information and contribute to $\beta$-component at an abstract level. 

The ranges of $\alpha$ and $\beta$ are (considering $TSS$ as even number):

\begin{equation}\label{4}
   \frac{2}{TSS} \leq \alpha \leq 1,
\end{equation} 

\begin{equation}
\begin{aligned}
    \frac{2}{H_p} \leq \beta & \leq 1, \quad  \text{if }  H_p \quad \text{is even, and} \\
    \frac{2H_p}{H_p^2 -1} \leq \beta & \leq 1, \quad \text{if }  H_p \quad \text{is odd}.
\end{aligned}\label{5}
\end{equation}

\begin{figure}[ht]
  \centering
  \includegraphics[width=0.8\linewidth,keepaspectratio]{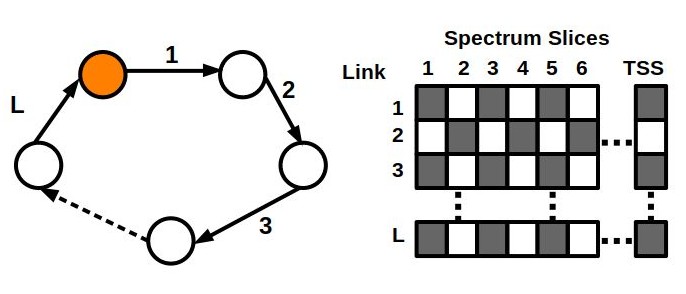}  
  \caption{Worst case fragmentation scenario in a network with $TSS$ number of spectrum slices, and continuity over a single path (an Euler path- covering all links of network without repetition, where number of visited nodes may repeat). }
  \label{fig:fig4}
\end{figure}

For the minimum value of $\beta$ for odd value of $H_p$, we can consider the chequered patterns as shown in Fig.\ref{fig:fig4}. In that case,

\begin{equation*}
\begin{aligned}
    \beta =  {1 \over 2} \left ( { 2 \over H_p -1 } + { 2 \over H_p +1} \right)
    = { 2 H_p \over H_p ^2 -1}.
\end{aligned} 
\end{equation*}

The lower limit of the range is calculated by considering a worst case scenario. In the worst case scenario each link in the network exhibits fig.~\ref{fig:fig3} case (d)'s spectrum status, with slices available alternatively in a link as well as in a path. As shown in the fig.~\ref{fig:fig4}, a single available slice is the maximum contiguous slot size in each of the links and also in the path(s). This scenario is further worsened by half of the available resources. We also get a highly fragmented scenario over a path, if maximum continuous slice over a single path is 1 for a spectrum index \textit{'i'} and half of the resources on that path are available. If such a case exists for all the spectrum slice indices, we get a worse case scenario for $\alpha$ as well as for the $\beta$. The best case scenario is when all the available slices/resources are available continuously and contiguously. Equations (\ref{4}) and (\ref{5}) present the range of $\alpha$ and $\beta$ respectively. The resultant of these components give $\nu$- or the vectored fragmentation metric (VFM) value in worst and best case scenarios.  

\subsection{An example of fragmentation level calculation}

\begin{figure}[t!]
    \begin{center}
        \includegraphics[width=0.8\linewidth,keepaspectratio]{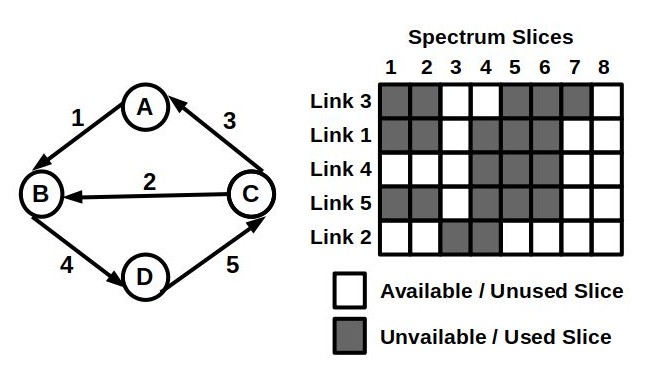}
        \caption{An example for calculation of vectored- fragmentation metric with 50 $\%$ occupancy}
        \label{fig:fig5}
    \end{center}
\end{figure}

In fig.~\ref{fig:fig5}, we have a 4-nodes 5-links network, with its network spectrum status. All the white blocks represent empty or available spectrum slices. In this spectrum scenario the occupancy level is 50$\%$. 

In the example, for the calculation of $\alpha$-component the average ratio of maximum contiguous slot size to total available spectrum slices is taken into account. In $\beta$-component calculation, a single longest path is taken into account which covers most of the source-destination pair routes. A single path traversing all of the links in 3-1-4-5-2 direction is considered for continuity calculation, hence $P$ = 1. The routes of nearly all the source-destination pairs are covered in this path. All the spectrum slice indices are having at least one available slice, in the given path, so $E_i$= 8.
In the example we calculate fragmentation level using Vectored fragmentation metric (AVFM), and compare the level with a link-based external fragmentation metric (L-EFM).\\
The L-EFM considers the largest contiguous slot's size ($CG_i$) in links, $i\in L$, to decide the fragmentation level.  It is the most basic metric with the least complexity. It also tends to ignore fragmentation due to smaller fragments present in the spectrum. The L-EFM formulation is given in equation \ref{6}.

\begin{equation} \label{6}
L-EFM= 1- \frac{\sum_{i=1}^{L} CG_i}{total available}
\end{equation}

The $\alpha$ and $\beta$ components for the above scenario are calculated as follows.
\begin{fleqn}
\begin{equation*}
    \alpha = \frac{1}{5}. (\frac{2}{3} + \frac{2}{3}+ \frac{3}{5} + \frac{2}{3} +\frac{4}{6}) = 0.6533 
\end{equation*}
\end{fleqn}

\begin{fleqn}
\begin{equation*}
    \beta =  \frac{1}{8} (\frac{1}{2} +\frac{1}{2}+ \frac{4}{4}+ \frac{1}{1}+ \frac{1}{1} +\frac{4}{4} +\frac{5}{5}) =  0.75 
\end{equation*}
\end{fleqn}

\begin{fleqn}
\begin{equation*}
    \nu = \sqrt{0.6533^2 + 0.75^2} = 0.9944
\end{equation*}
\end{fleqn}

\begin{fleqn}
\begin{equation*}
   VFM_{min}= \nu _{min} = 0.486 
\end{equation*}
\end{fleqn}

Normalized Vectored Fragmentation Metric (NVFM),

\begin{fleqn}
\begin{equation*}
   NVFM = \nu _{norm} = \frac {\nu- \nu _{min}}{\nu_{max} - \nu_{min}} = \frac{0.9944 -0.486 }{1.414 - 0.486} = 0.547844  
\end{equation*}
\end{fleqn}

\begin{fleqn}
\begin{equation*}
   AVFM= 1-\nu _{norm} = 0.452156 
\end{equation*}
\end{fleqn}

\begin{fleqn}
\begin{equation*}
    L-EFM=1- \frac{2+2+3+2+4}{20} = 1- \frac{13}{20} = 0.35
\end{equation*}
\end{fleqn}

\medskip
In this particular example, the network spectrum utilization is 50$\%$, that is half of the total wavebands are currently used. The corresponding normalized vectored fragmentation metric (NVFM) value is around 0.547844. This means that a moderate fragmentation is present in the spectrum. If the $\nu$ (VFM) value is closer to $\sqrt{2}$ (or $\nu _ {norm}$ (NVFM) is closer to 1), then fragmentation in the network spectrum is not significant. While comparing the VFM with other link-based fragmentation metrics, we use the adapted form of VFM (AVFM), i.e., $(1- \nu_{norm})$. This ensures that same meaning is conveyed effectively, that is low metric value means lower level of fragmentation and vice versa. A direct relationship between the network spectrum utilization and the fragmentation level as vectored fragmentation metric is very unlikely, as the way in which the spectrum is occupied also plays a part. The fragmentation is lowest when there is small spectrum utilization and large spectrum utilization. For the midway scenario, the metric essentially depends on the state of network spectrum, and the relation between adjacent slices, not on the total number of slices. In the real time traffic scenario, the fragmentation level can vary for the same network spectrum utilization, when observed in steady state condition. 
\begin{table*}[t!]
\caption{Comparison of fragmentation metrics}
\begin{center}
\resizebox{12cm}{!}{
\begin{tabular}{|c|c|c|c|}
\hline

\textbf{Characteristics} & \textbf{\textit{Link- based}}& \textbf{\textit{Path- based}}& \textbf{\textit{Vectored}} \\
\hline
\hline
Identifies & & &  \\
fully fragmented scenario & No & No & Yes   \\
(Fig.3(d)) & & &   \\
\hline
Identifies & & &  \\
zero fragmentation scenario & Yes & Yes & Yes \\
(Fig.3(a,b,c)) & & &   \\
\hline
Differentiates between & &  &   \\
fragmentation scenarios & Some of them & Yes & Yes   \\
(Fig. 3(e)) & & &   \\
\hline
                & Lowest among others & High & Moderate \\
  Time Complexity & $\mathcal{O}(S.L)$ \cite{c10} & $\mathcal{O}(S.L)$-$\mathcal{O}(S.L.G)$ \cite{c10} & $\mathcal{O}(2.S.L)$\\
 & &  &  \\
\hline
General Observations
&
\parbox{4.0cm}{\begin{flushleft} 1. Ignores small fragments\\
2. Can be relative or absolute\\
3. Covers only contiguity aspect\\
\end{flushleft}
}
&
\parbox{4.0cm}{
\begin{flushleft}1. Specific to path of the connection request\\
2. Relative\\
3. Covers contiguity and continuity aspect specific to connection request\\
\end{flushleft}
}
&
\parbox{4.0cm}{
\begin{flushleft}
1. Covers all available spectrum slices\\
2. Absolute\\
3. Covers both contiguity and continuity aspect independently.\\
\end{flushleft}
}\\

\hline
\end{tabular}
\label{tab:tab2}
}
\\\begin{flushleft}
\footnotesize{S is total number of spectrum slices, L is total number of Links and G is total number of permissible granularity of connection requests.}\\
\end{flushleft}
\end{center}
\end{table*}

We compare theoretical aspects of the vectored fragmentation metric and the other metrics reported in the literature. The comparison allows us to put vectored fragmentation metric in perspective. The comparison is based on some of the essential characteristics as well as the complexity of the metrics. Table \ref{tab:tab2} compares link-based metric, path-based metric and the vectored metric using some key features such as the ability to identify fragmentation scenario and the time complexity. The vectored fragmentation metric can outperform link-based fragmentation metrics for fragmentation estimation at the network level with an additional computation cost which is still less than that of any path-based fragmentation metric.

In the simulation study of fragmentation indicator AVFM, we compare it with only link based metric, L-EFM. The L-EFM is the simplest representation of fragmentation level and does not have any pre-conditions. The L-EFM covers only the contiguity aspect, and thus, comparison with a continuity based metric is pertinent in fragmentation study. 

\subsection{Evaluation of the vectored fragmentation metric}

\begin{enumerate}
\item $\alpha$-component presents the consolidated fragmentation of individual links, covering the fragmentation due to contiguity constraint.
\item $\beta$-component presents the consolidated fragmentation on a single slice index over a multi-hop path, covering the fragmentation due to continuity constraint.
\item Both $\alpha$ and $\beta$-component can be accepted as a measure of fragmentation individually.
\item The connectivity of multiple slices over a path in $\beta$-component is ignored here. We take the continuity and contiguity as independent entities here.
\item Higher is the value of the vectored fragmentation metric ($\nu$ or $\nu_{norm}$), lower is the fragmentation in the network spectrum.
\item This metric is independent of any particular connection request type. 
\item This is $1^{st}$ level metric, as it takes into account only single slice index for the evaluation of continuity constraint fragmentation. 
\item When $\beta$-component is individually taken as a fragmentation indicator, and we check continuity of  $n$ contiguous slices over a path, it becomes a $n^{th}$-level metric.
\item The computational complexity in finding fragmentation level in a network with \textit{N} nodes, \textit{L} links and \textit{S} spectrum slices on each link, using Vectored fragmentation metric is $\mathcal{O}(2.S.L)$.
\end{enumerate}

\section{Results and Discussions}

The table \ref{tab:tab2} presents a qualitative comparison between the link-based, path-based, and vectored fragmentation metric. To study the metrics quantitatively, we use the link-based and vectored fragmentation metric in a real-time traffic scenario in some example networks.  We consider three networks:  Net-A with 7-nodes 12-bidirectional links, NSF network with 14-nodes 21-bidirectional links, and German network with 17-nodes 26-bidirectional links (table ~\ref{tab:net}). We generate connection requests, also called demands, on every node using Poisson distributed arrival process and exponentially distributed holding time. At each node, the arrived connection requests' destinations are selected with equal probability from the other nodes. We consider a randomly distributed bandwidth requirement (in terms of spectrum slices) ranging from one to some maximum permissible slices for each connection request (including guard band). In the networks, each link is considered to have 320 spectrum slices. We assume no waveband conversion scheme at any intermediate node. Dijkstra's shortest path algorithm is used to find a route between each source-destination pair. Thereafter, the spectrum slice assignment is done according to the first-fit (FF) strategy. In case, a connection request cannot be satisfied, it is dropped. We obtain all the results with a 99$\%$ confidence interval by averaging over multiple simulation runs.

We analyze our fragmentation metric for steady-state and transient-state scenarios for constant load condition. The transient-state starts from the initial empty network spectrum state. The spectrum state changes with the arrivals and departures of connection requests and eventually settles to a steady-state. There is not much variation with time in the characteristics of the network spectrum in steady-state. The steady-state observations indicate normal operating status. It does not give any insight into how the fragmentation shapes up in the spectrum with time. To analyze how fragmentation level changes with changing network state dynamics, we need to evaluate the fragmentation metric and other parameters in the transient state.

\begin{table}[!t]
\caption{Network Scenarios for Evaluation of Fragmentation Metrics in EONs}
\label{tab:net}
\centering
\resizebox{0.6\columnwidth}{!}{
\begin{tabular}{|c|c|c|c|}
\hline

\textbf{Properties} & \textbf{\textit{NET-A}}& \textbf{\textit{NSFNET}}& \parbox{2.5cm}{\textbf{\textit{GERMAN NET}}} \\
\hline
\parbox{2.2cm}{Nodes} & 7 & 14 & 17  \\
\hline
\parbox{2.2cm}{Links} & 24 & 42 & 56   \\
\hline
\parbox{2.2cm}{\begin{flushleft}Average Nodal degree\end{flushleft}} & 3.42& 3 & 3.05  \\
\hline
\parbox{2.2cm}{\begin{flushleft}Paths, P in eqn.(2)\end{flushleft}} & 2 & 10 & 6   \\
\hline
\end{tabular}
}
\label{tab:tab3}
\end{table}

\subsection{Simulation Results}
\begin{figure}[!t]
\begin{subfigure}{.45\linewidth}
  \centering
  \includegraphics[width=\textwidth]{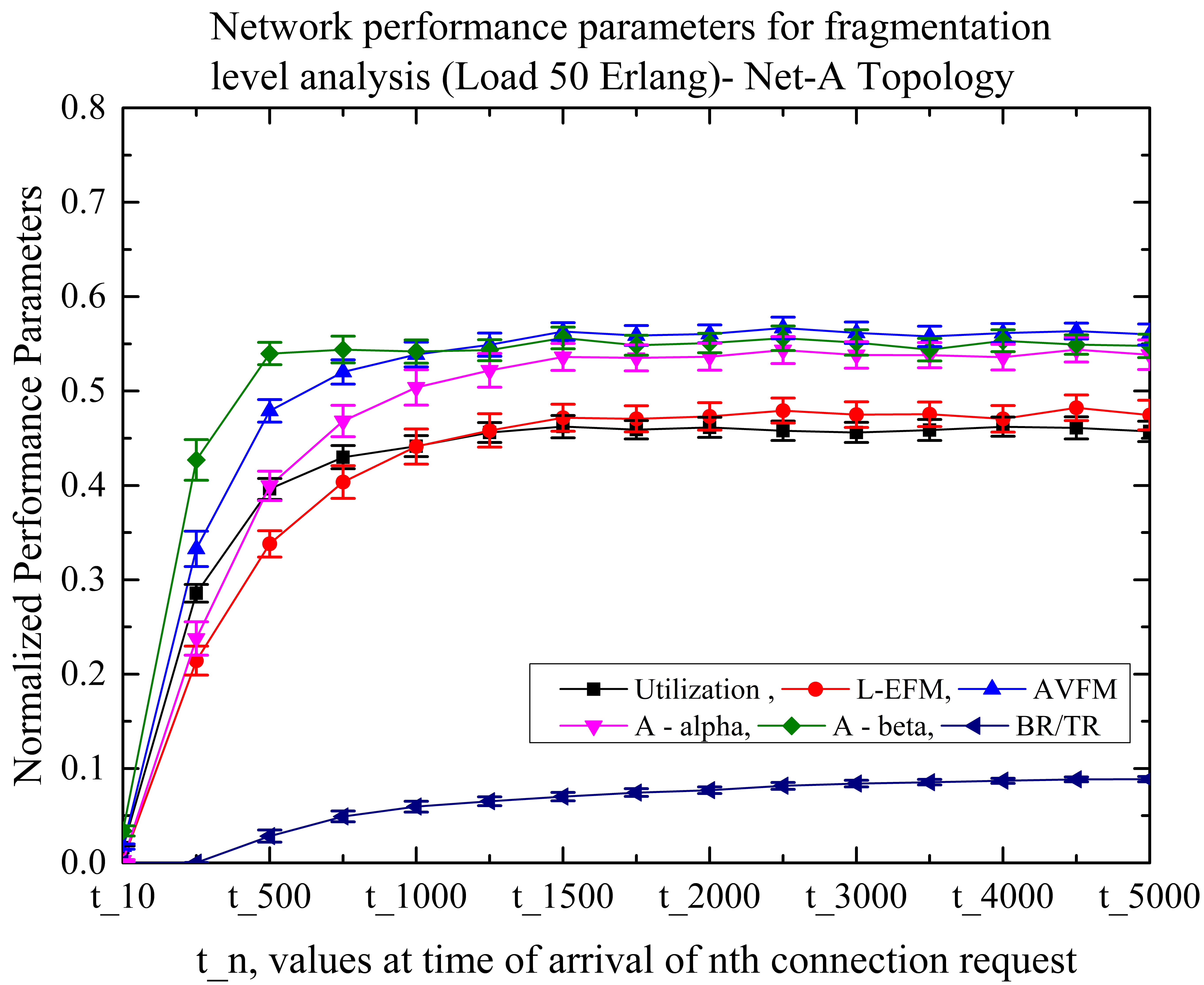}  
  \caption{}
  \label{fig:FigNetA1}
\end{subfigure}
\begin{subfigure}{.45\linewidth}
  \centering
  \includegraphics[width=\textwidth]{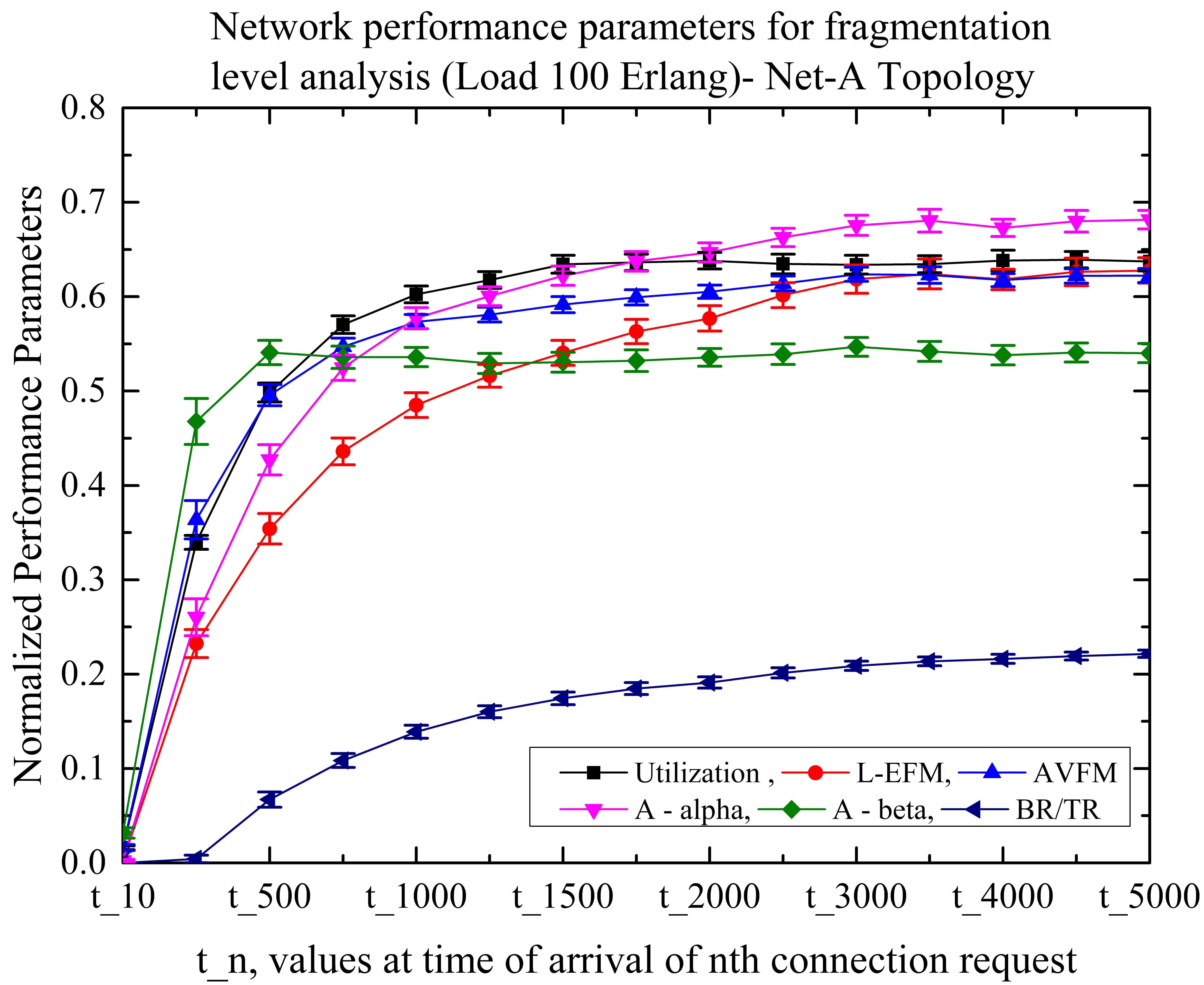}  
  \caption{}
  \label{fig:FigNetA3}
\end{subfigure}
\caption{Evolution of fragmentation level with time for traffic loads (a) 50 E, and (b) 100 E in Net-A Network topology}
\label{fig:fig6}
\end{figure}

\begin{figure}[!t]
\begin{subfigure}{.45\linewidth}
  \centering
  \includegraphics[width=\textwidth]{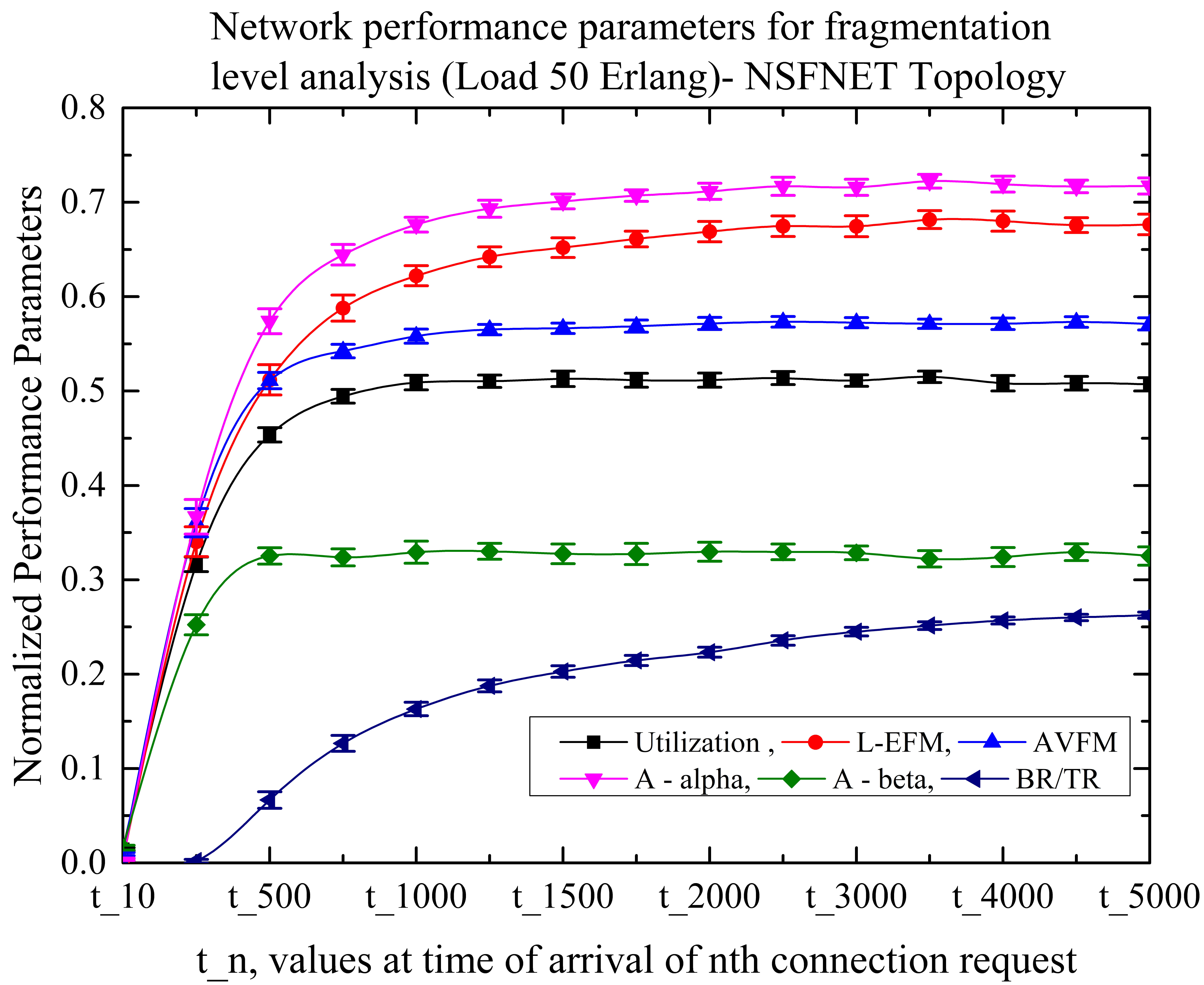}  
  \caption{}
  \label{fig:FigNSF1}
\end{subfigure}
\begin{subfigure}{.45\linewidth}
  \centering
  \includegraphics[width=\textwidth]{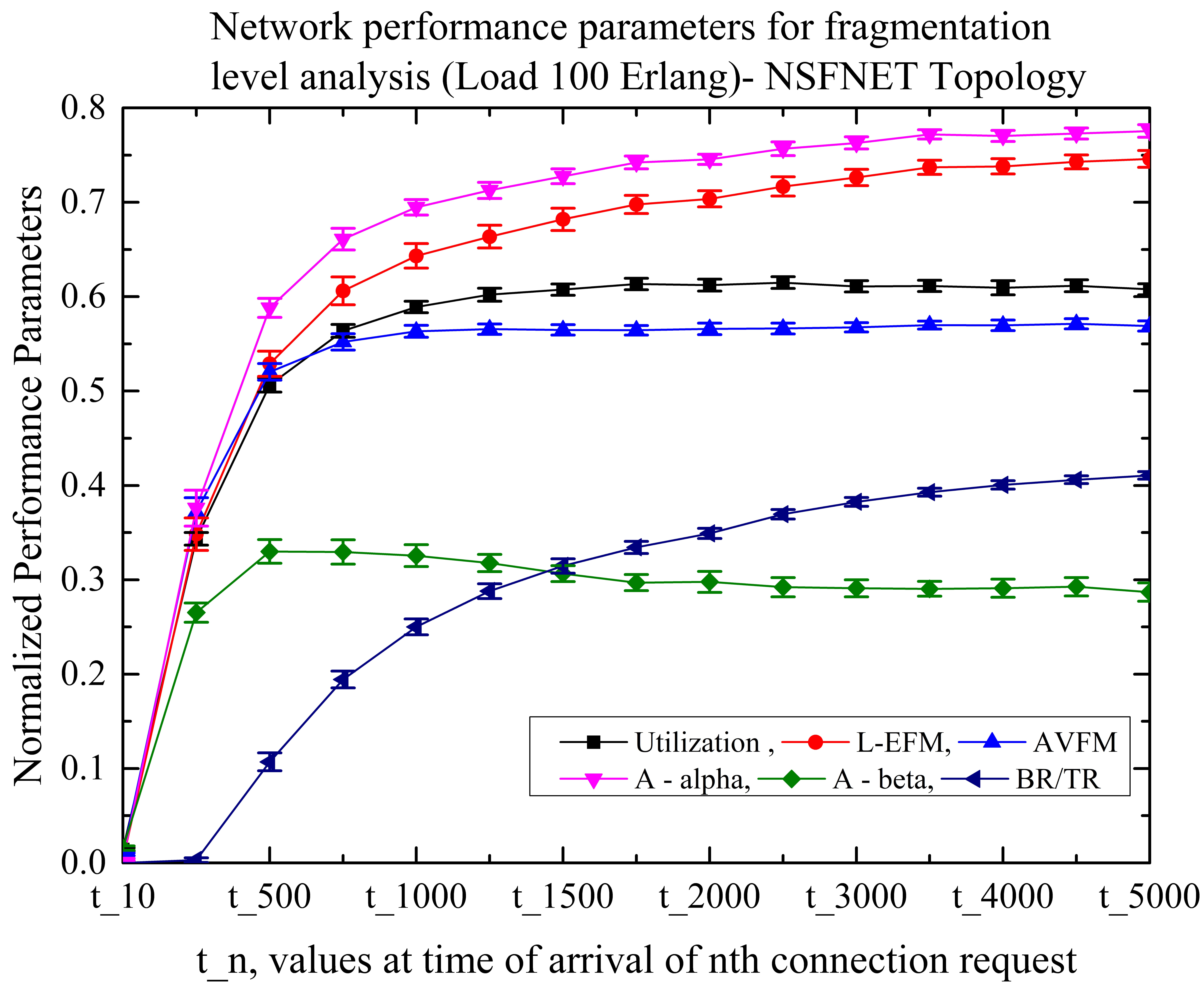}  
  \caption{}
  \label{fig:FigNSF3}
\end{subfigure}
\caption{Evolution of fragmentation level with time for traffic loads (a) 50 E, and (b) 100 E in NSF Network topology}
\label{fig:fig7}
\end{figure}

\begin{figure}[!t]
\begin{subfigure}{.45\linewidth}
  \centering
  \includegraphics[width=\textwidth]{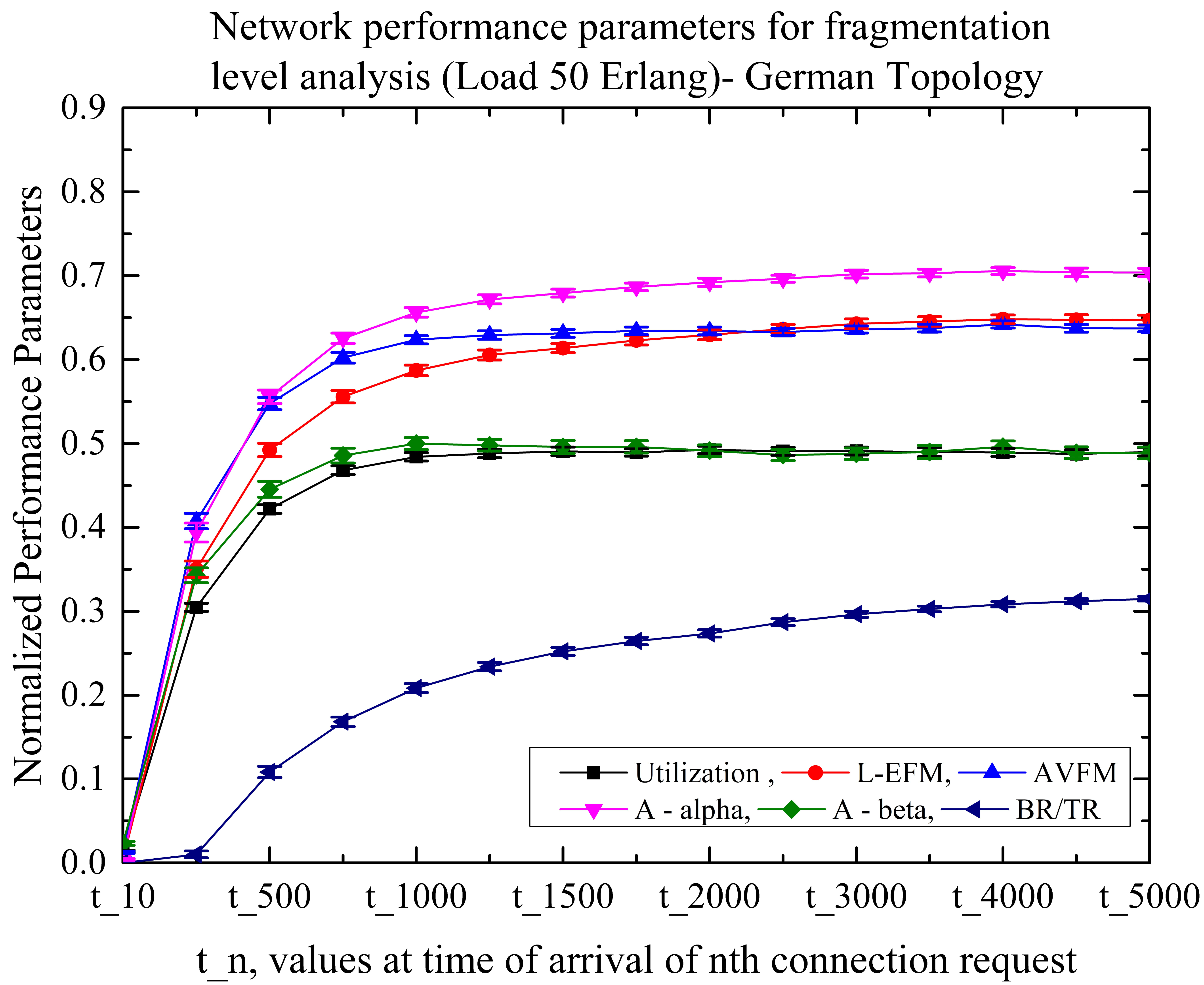}  
  \caption{}
  \label{fig:FigGER1}
\end{subfigure}
\begin{subfigure}{.45\linewidth}
  \centering
  \includegraphics[width=\textwidth]{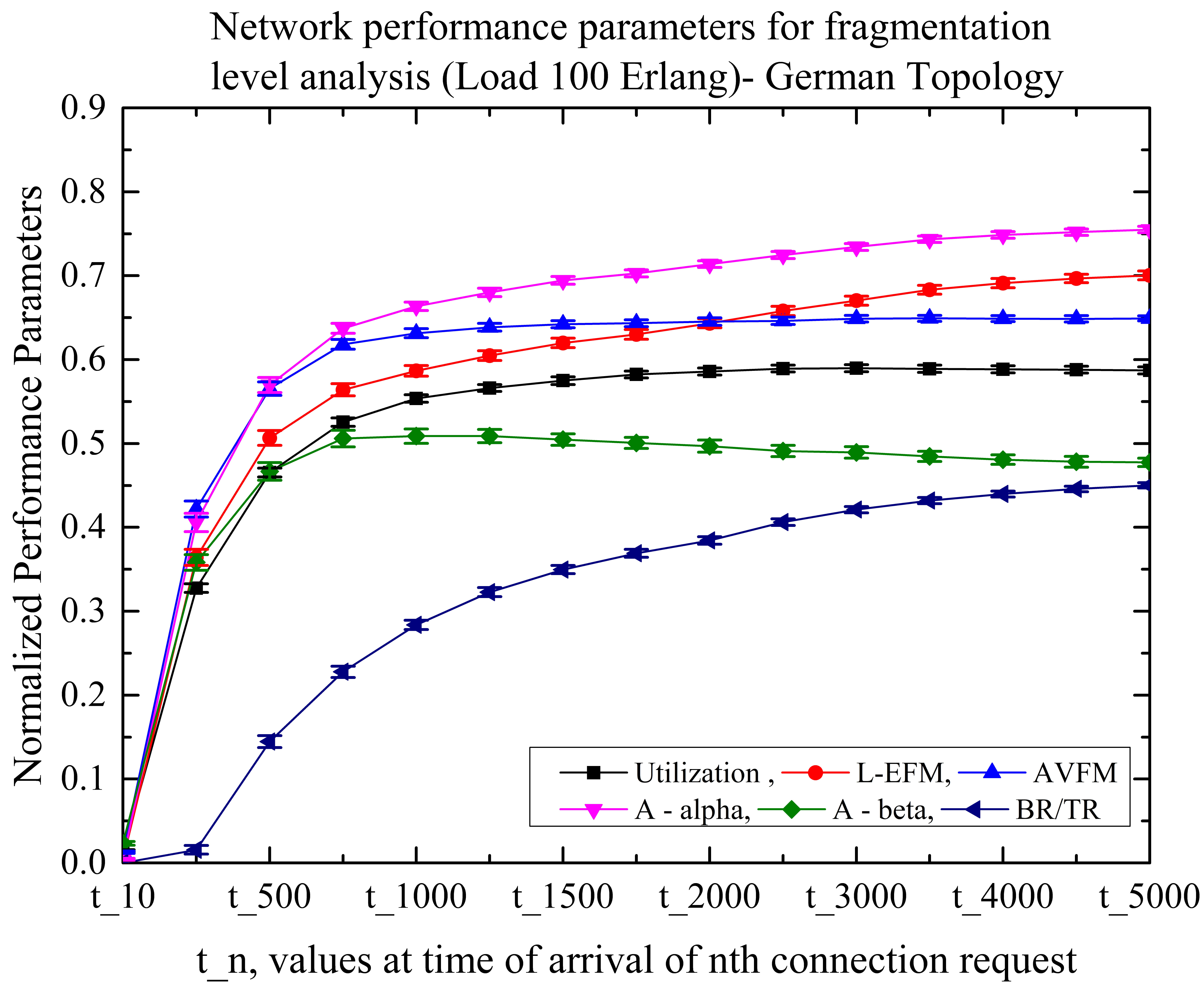}  
  \caption{}
  \label{fig:FigGER3}
\end{subfigure}
\caption{Evolution of fragmentation level with time for traffic loads (a) 50 E, and (b) 100 E in German Network topology}
\label{fig:fig8}
\end{figure}

\subsubsection{Transient-State Scenario}

We observed the transient state parameters to understand their time evolution for three networks and two traffic load\footnote{load or average load means traffic load at every node of the network, wherever mentioned} conditions of 50 Erlangs (low load) and 100 Erlangs (high load). The maximum permissible slice requirement by a connection request is sixteen. We started from a completely available/ empty network. The connection requests arrive, set up, and finally released dynamically on the completion of their holding time. Some of them are blocked if set up is not feasible. We observed the evolution of average network spectrum utilization (Utilization in figs.~\ref{fig:fig6}, \ref{fig:fig7}, \ref{fig:fig8}), the average link-based external fragmentation metric (L-EFM), the average blocked requests to total requests ratio (BR/TR), and the average adapted vectored-fragmentation metric (AVFM), i.e., (1-$\nu_{norm}$), till the arrival of initial 5000 connection requests. We also observed average adapted-$\alpha$ (A-alpha) and adapted-$\beta$ (A-beta) components as individual fragmentation indicators as defined by relations in eq.\eqref{1} and eq.\eqref{2}'s adapted form. All the observed parameters increase initially with the incoming connection requests and then attain a steady-state value without any oscillatory behaviour, in all the graphs. 

For Net-A, at low load condition (fig.\ref{fig:fig6}(a)), A-beta (continuity fragmentation) increases faster and remains slightly higher than A-alpha (contiguity fragmentation). For higher load conditions (fig.\ref{fig:fig6}(b)) also, A-beta increases faster and attains a steady state value. A-alpha increases relatively slower but exceeds the A-beta. This indicates at high load conditions, contiguity fragmentation is dominant factor. AVFM values always lie in between A-alpha and A-beta. L-EFM value is lower than AVFM but follows the same pattern. It can be also seen that L-EFM is lower than AVFM initially, but it becomes nearly equal to AVFM in steady-state. The network utilization and BR/TR ratio also increases with time to attain a steady-state.

For NSFNET and German network, we observed that A-alpha parameter changes at a faster rate and is the major contributor to fragmentation level (fig.\ref{fig:fig7}(a) and (b), fig.\ref{fig:fig8}(a) and (b)) for both the loading conditions. Here, in fig.\ref{fig:fig7}(a) and (b) for NSFNET, the L-EFM is greater than AVFM value and it follows A-alpha (both being contiguity component) instead of AVFM. In fig.\ref{fig:fig8}(a) and (b), for German network also, L-EFM follows same pattern as A-alpha. L-EFM is lower than AVFM for some initial time and then crosses over the AVFM value. The reason could be contribution of A-beta and number of paths in AVFM. A-beta increases and attains steady-state value quickly. In NSFNET and German network topology, we consider multiple paths for A-beta calculation. The fragmentation on these paths is relatively low as the spectrum slices' unavailability on a path is considered no fragmentation. A-beta increases initially and then is seen to slightly fall. The reason could be busy network resources themselves on the selected paths. 

In the time evolution plots for the three network topologies, fragmentation level indicators, AVFM and L-EFM develop gradually as expected. The transient state evolution of the AVFM allows us to reach a conclusion that A-alpha and A-beta together can be used to check the fragmentation level, considering some pre-defined parameters (like number of paths and number of spectrum slices). Next, We observe the proposed AVFM's performance for varying network loading conditions, starting with A-alpha and A-beta dominance study. A-alpha dominates A-beta throughout the observation time for both load conditions in NSFNET and German network. So, the network conditions affecting dominance of A-alpha and A-beta are also important.

In A-beta, if the number of considered paths is large, then the fragmentation due to continuity is not significant, and the output is merely the availability of spectrum resources on a path. So, the A-beta may not be of significance depending on whether the static or dynamic routing conditions are considered as discussed in section III A. 

The A-alpha and L-EFM are same in pattern, only their values are different. L-EFM can be replaced by A-alpha as link fragmentation indicator for smaller networks or networks with multiple paths of shorter lengths. The smaller networks will have paths with shorter length.


\subsubsection{Steady-State Scenario}
In a transient-state, we established that AVFM in the network could follow fragmentation level development, just like L-EFM. But, there may be more awareness of the two contributors, A-alpha and A-beta, in AVFM. Next, we studied AVFM elaborately in the steady-state. We first studied performance measures of A-alpha and A-beta over varying load conditions. We also studied the performance measure of AVFM for different connection requests' arrival rates, holding times, and the maximum permissible granularity range\footnote{ labeled MaxDemand in related plots}.  
\begin{figure*}
\begin{subfigure}{.34\textwidth}
  \centering
  \includegraphics[width=.7\linewidth]{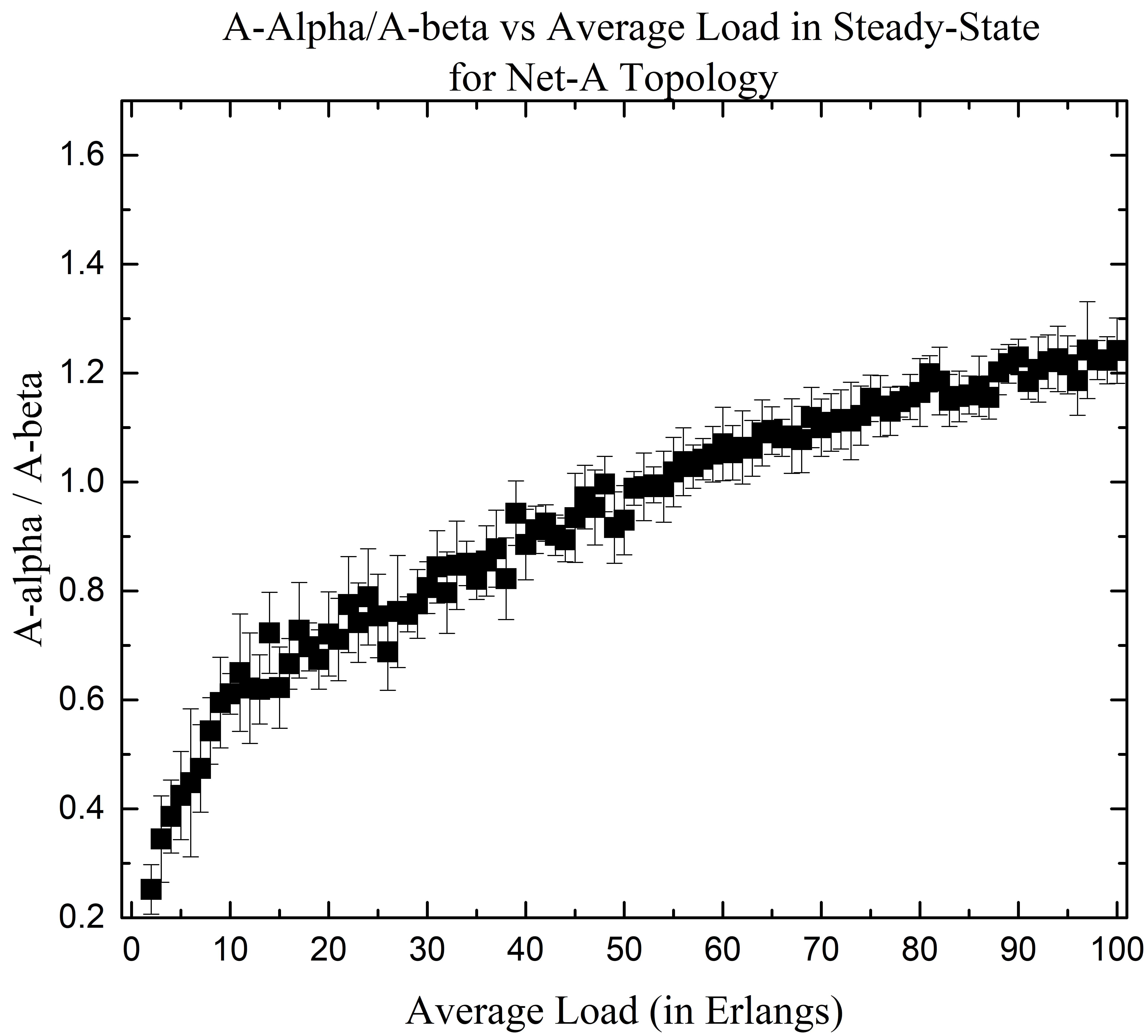}  
  \caption{}
  \label{fig:HTNetA}
\end{subfigure}
\begin{subfigure}{.34\textwidth}
  \centering
  \includegraphics[width=.7\linewidth]{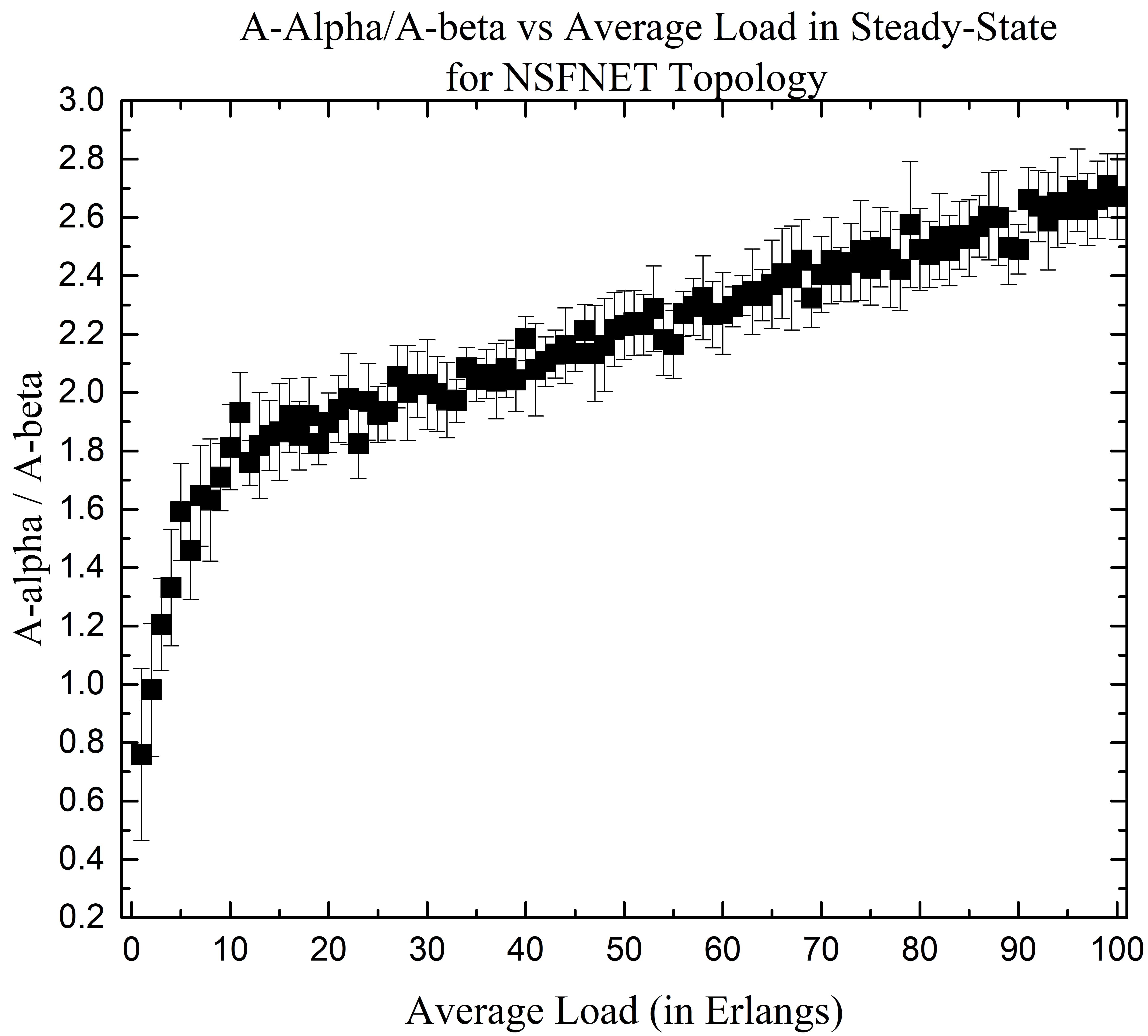}  
  \caption{}
  \label{fig:HTNSF}
\end{subfigure}
\begin{subfigure}{.34\textwidth}
  \centering
  \includegraphics[width=.7\linewidth]{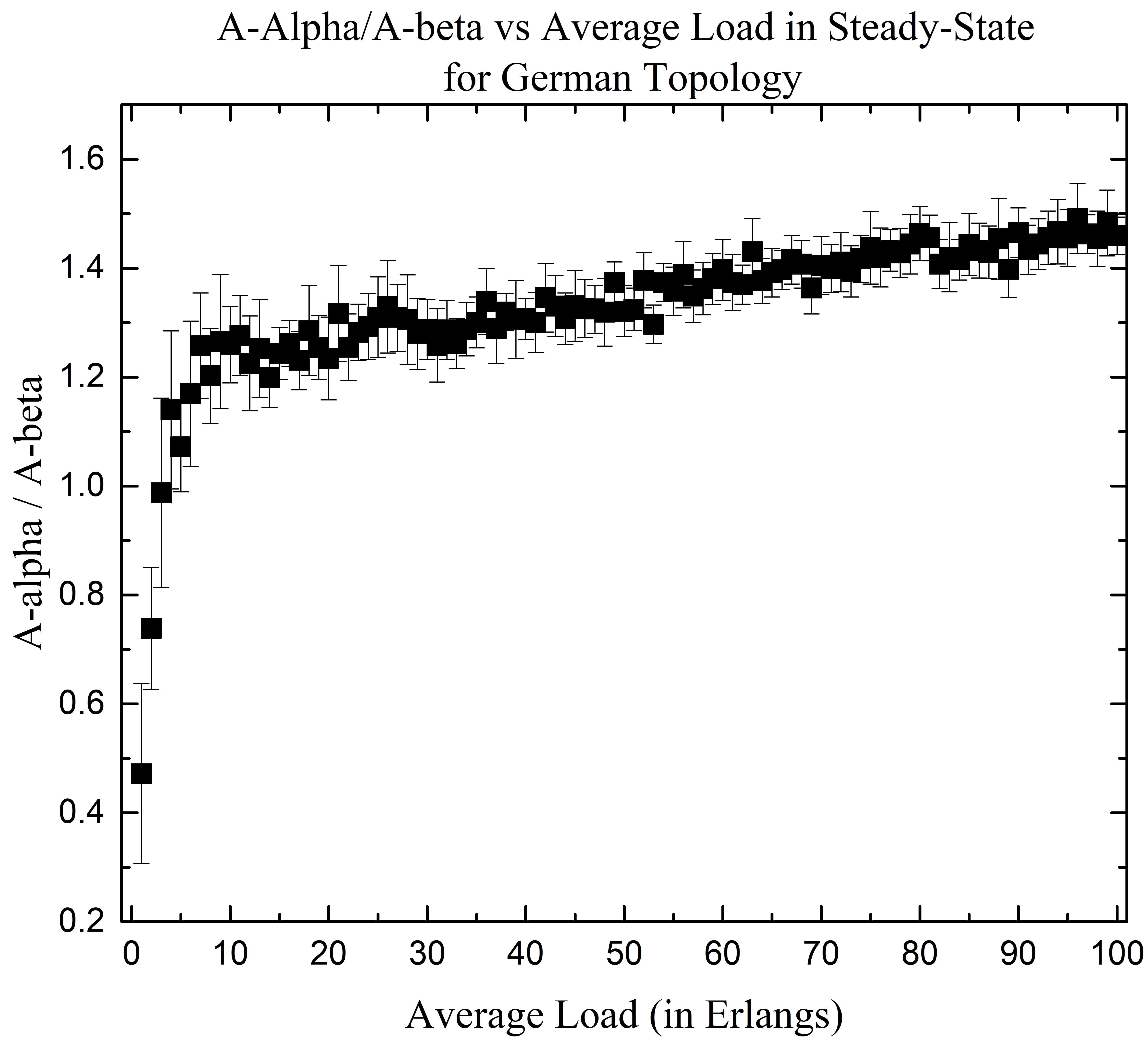}  
  \caption{}
  \label{fig:HTGER}
\end{subfigure}
\caption{Ratio of A-alpha and A-beta vs traffic load in steady-state for (a) Net-A, (b) NSFNet, and (c) German topology for different average holding times.}
\label{fig:fig9}
\end{figure*}

\begin{figure*}
\begin{subfigure}{.34\textwidth}
  \centering
  \includegraphics[width=.7\linewidth]{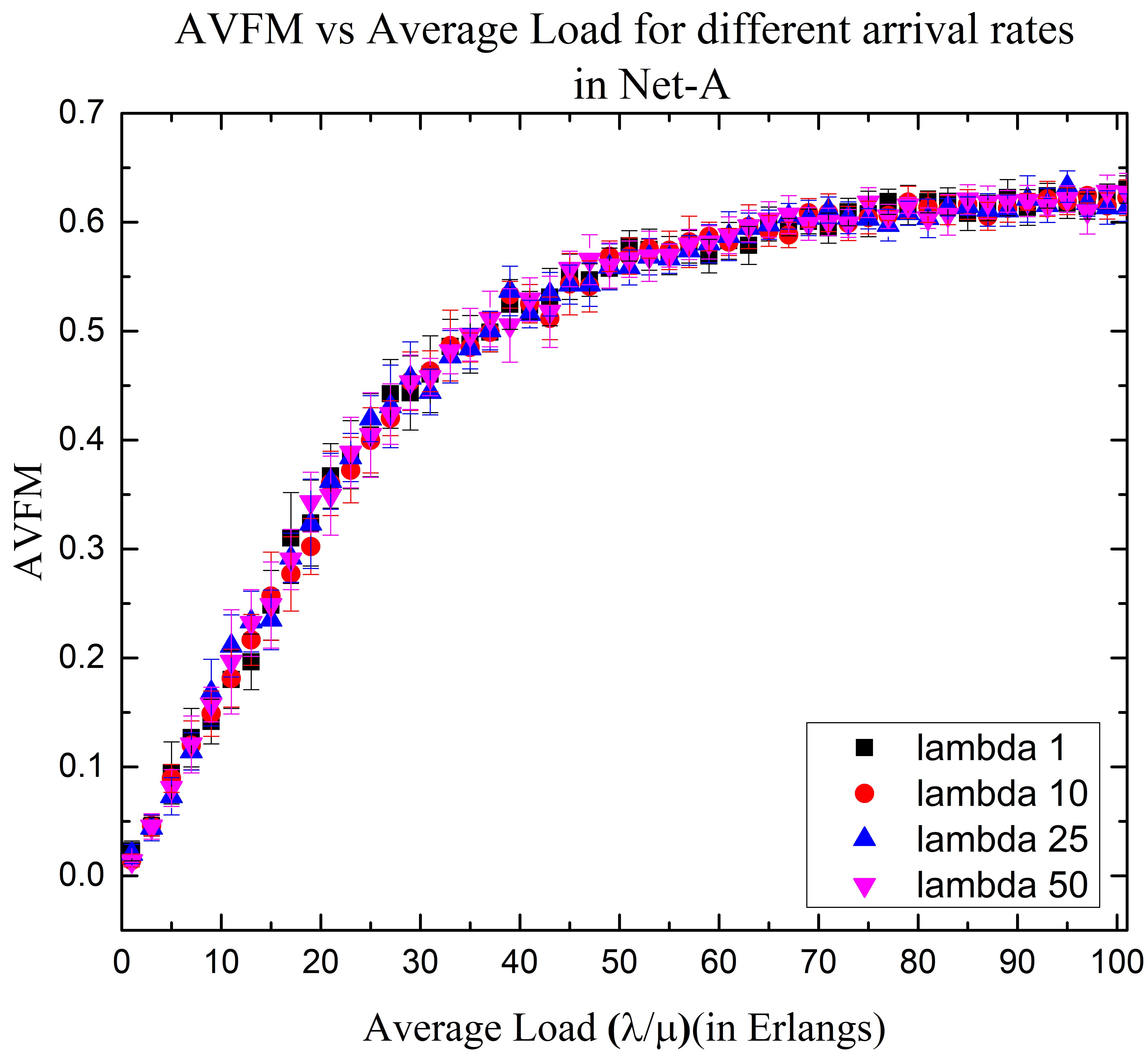}  
  \caption{}
  \label{fig:LoadNetA}
\end{subfigure}
\begin{subfigure}{.34\textwidth}
  \centering
  \includegraphics[width=.7\linewidth]{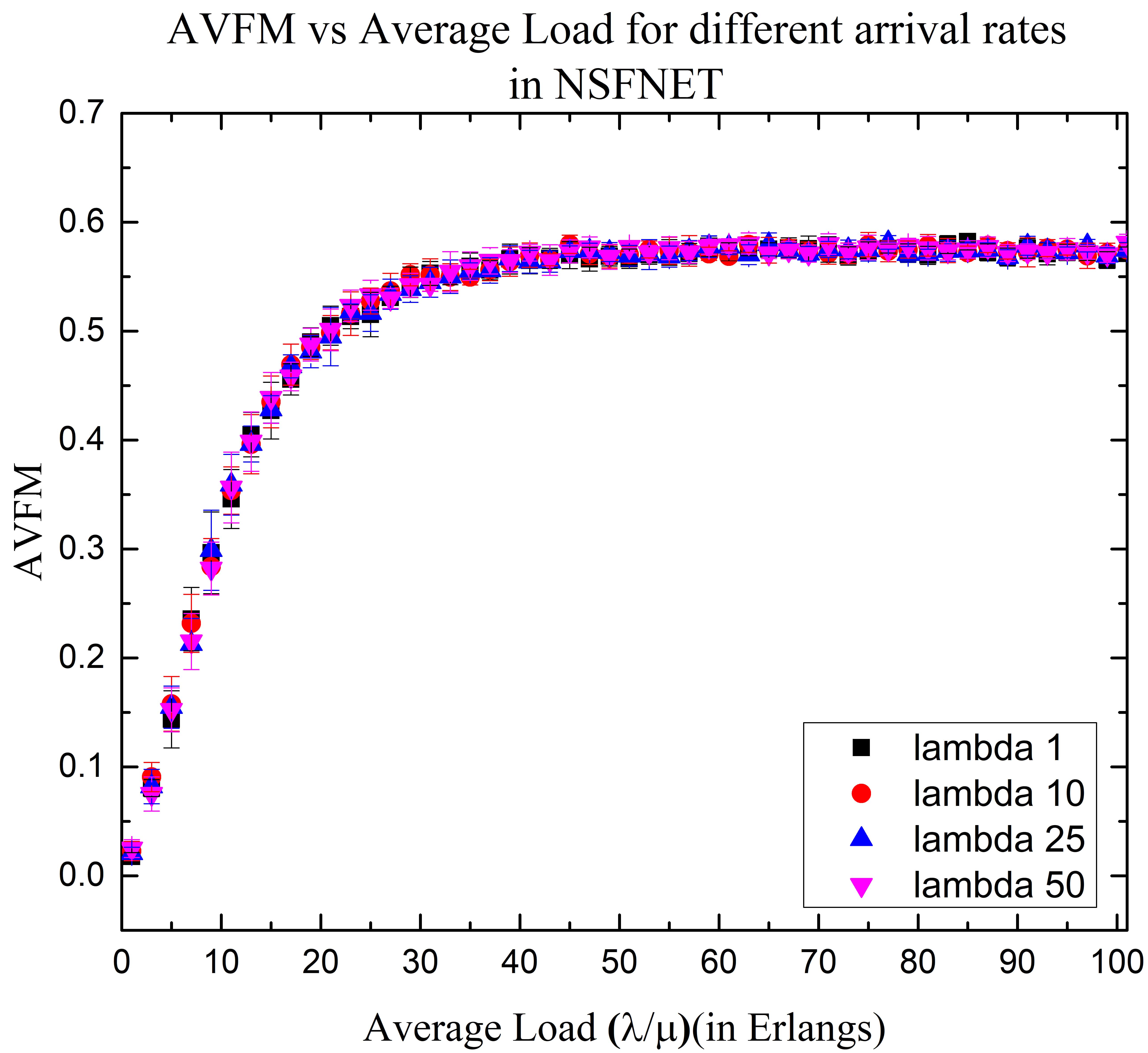}  
  \caption{}
  \label{fig:HTNSF}
\end{subfigure}
\begin{subfigure}{.34\textwidth}
  \centering
  \includegraphics[width=.7\linewidth]{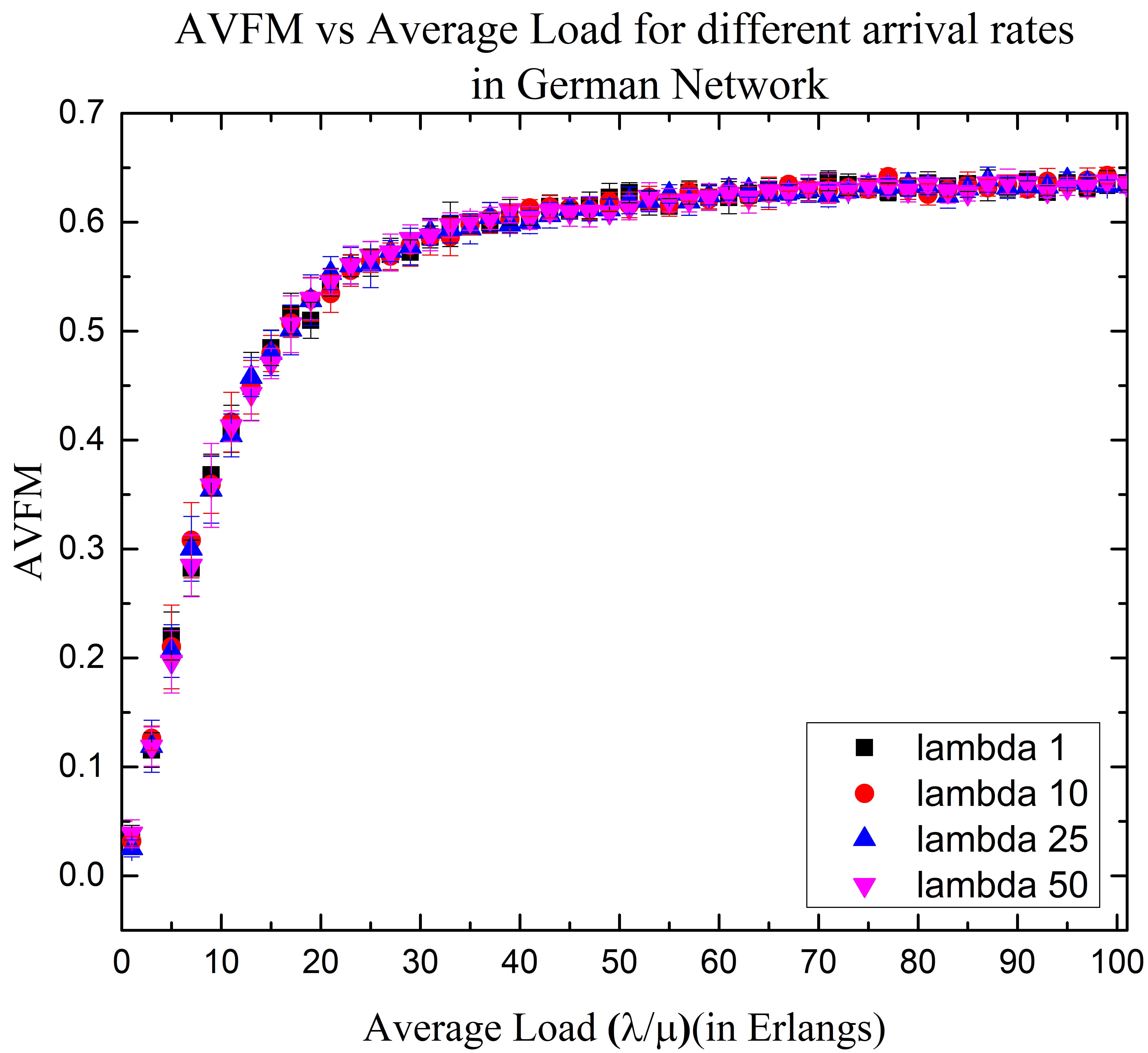}  
  \caption{}
  \label{fig:LoadGER}
\end{subfigure}
\caption{AVFM vs traffic load in steady-state for (a) Net-A, (b) NSFNET, and (c) German topology for different arrival rates (lambda).}
\label{fig:fig10}
\end{figure*}

In fig.\ref{fig:fig9}(a)-(c), we observe how the influence of A-alpha and the A-beta as fragmentation level indicator vary for different network topologies. In Net-A, at low load traffic, A-beta is dominant than the A-alpha (A-alpha/A-beta $<$ 1) for traffic load of 1 Erlang to 50 Erlangs. As already discussed, the A-beta influences the Net-A more due to fewer path and longer path length consideration. The crossover point is when A-alpha becomes more than A-beta.  For Net-A, the crossover point occurs at around 52 Erlangs. This crossover point arrives at low load ($\leq 10 Erlangs$) in NSFNET and German network because there are many paths for continuity test, and the major contributor in them is contiguity fragmentation on individual links. It can be seen in fig.\ref{fig:fig9}(b)-(c) that A-alpha is the dominant component for a wide range of the average load at nodes. It again implies that the A-alpha is sufficient to track fragmentation for larger networks, where a more number of the paths are needed for A-beta. 

There is one inconsistency in A-beta's contribution also.  Here, we check the continuity of the individual spectrum index, and there may be a single slice available on each link of the path but at the different indices. Then this type of fragmentation goes unnoticed by the metric. However, there is very little likelihood of this event, as it requires a very high network utilization value (almost full use of network spectrum). This situation can only arise when there may be non-uniform/irregular traffic distribution, creating a crunch of available capacity. Also, there is no requirement of A-beta in a full mesh network topology with fixed routing. There will be single link paths between all source-destination pairs, and the A-alpha component is sufficient to evaluate the fragmentation level.

Fragmentation level depends on the network spectrum status which in turn depends on the network traffic dynamics. The varying traffic dynamics could be the average arrival rate, the average holding time, or the MaxDemand (acceptable granularity) of connection requests. In fig.\ref{fig:fig10}(a)-(c), we considered four different arrival rates (lambda 1, lambda 10, lambda 25, and lambda 50), each one with increasing holding times to create varying load conditions. We observed that AVFM increases with the traffic load. The maximum steady-state fragmentation level in Net-A is around 0.6, i.e., 60 percent fragmentation, for traffic load up to 100 Erlangs. In NSFNET and German network (fig.\ref{fig:fig10}(b)-(c)), the steady-state AVFM value varies from 0.55 to 0.6 for an average load of 40 Erlangs till 100 Erlangs. We observe that the steady-state AVFM value does not change for a fixed load. So, varying the arrival rate and holding time for a fixed load will result in the same fragmentation level.



\begin{figure*}
\begin{subfigure}{.34\textwidth}
  \centering
  \includegraphics[width=.7\linewidth]{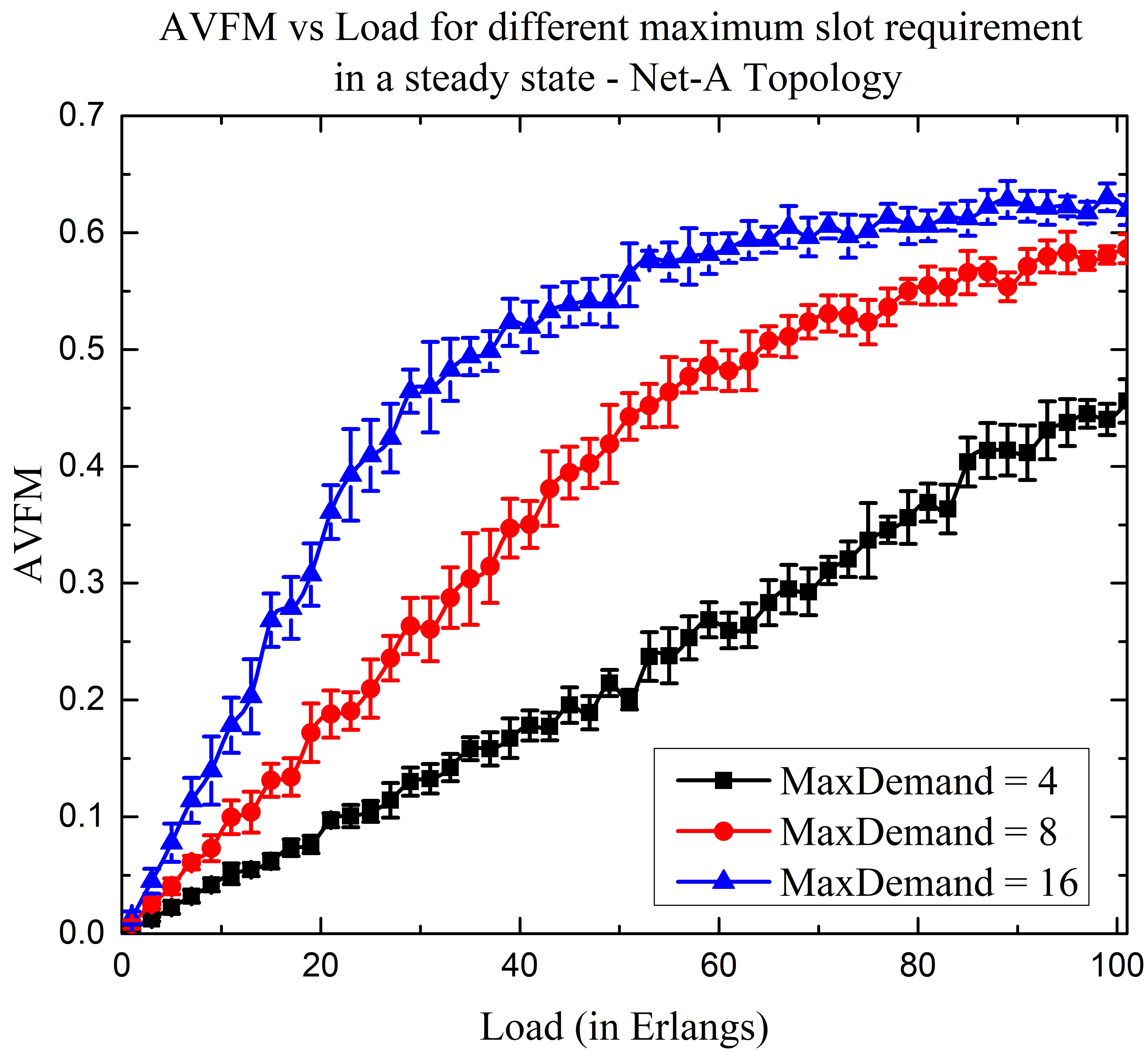}  
  \caption{}
  \label{fig:DMNetA}
\end{subfigure}
\begin{subfigure}{.34\textwidth}
  \centering
  \includegraphics[width=.7\linewidth]{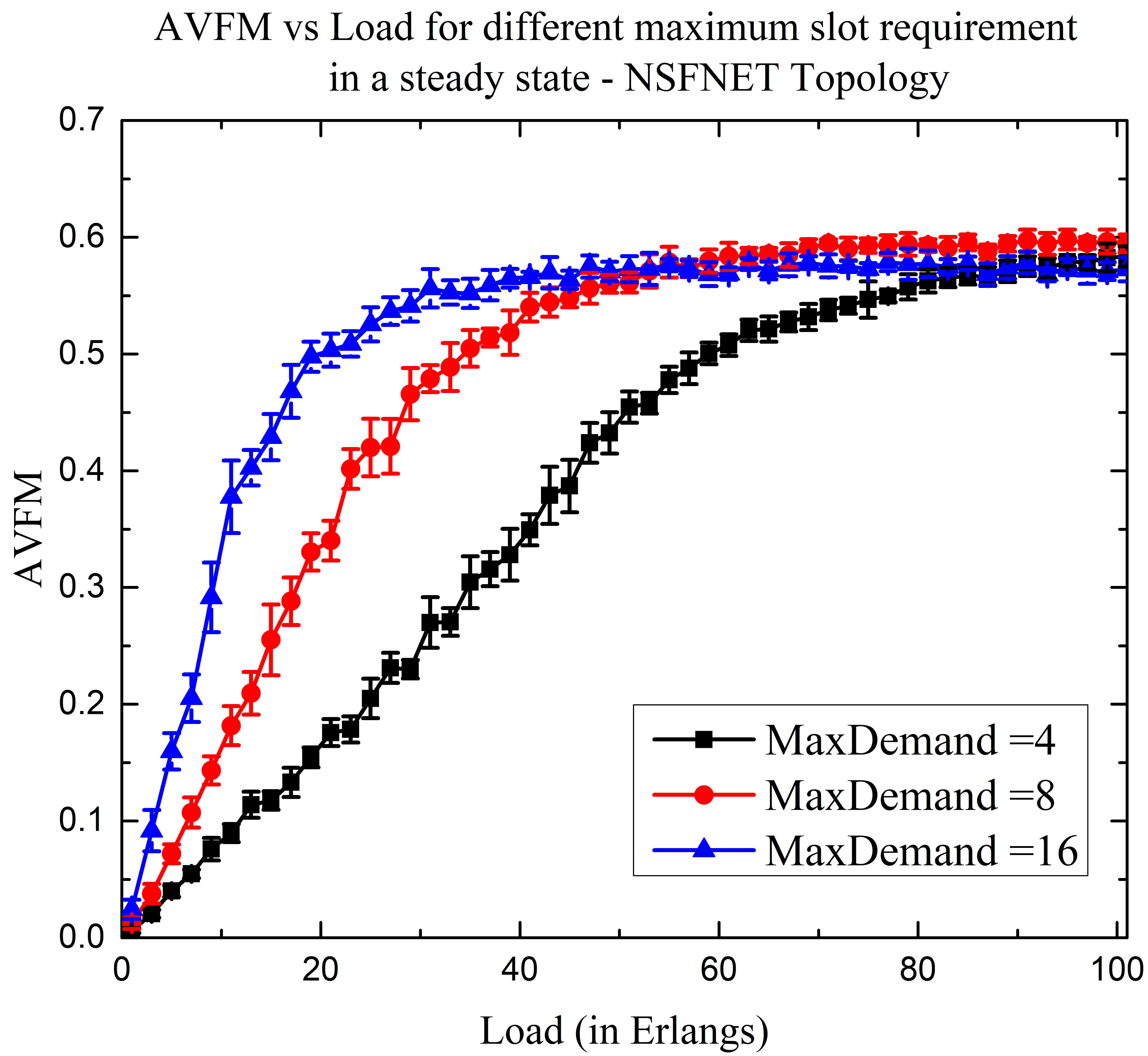}  
  \caption{}
  \label{fig:DMNSF}
\end{subfigure}
\begin{subfigure}{.34\textwidth}
  \centering
  \includegraphics[width=.7\linewidth]{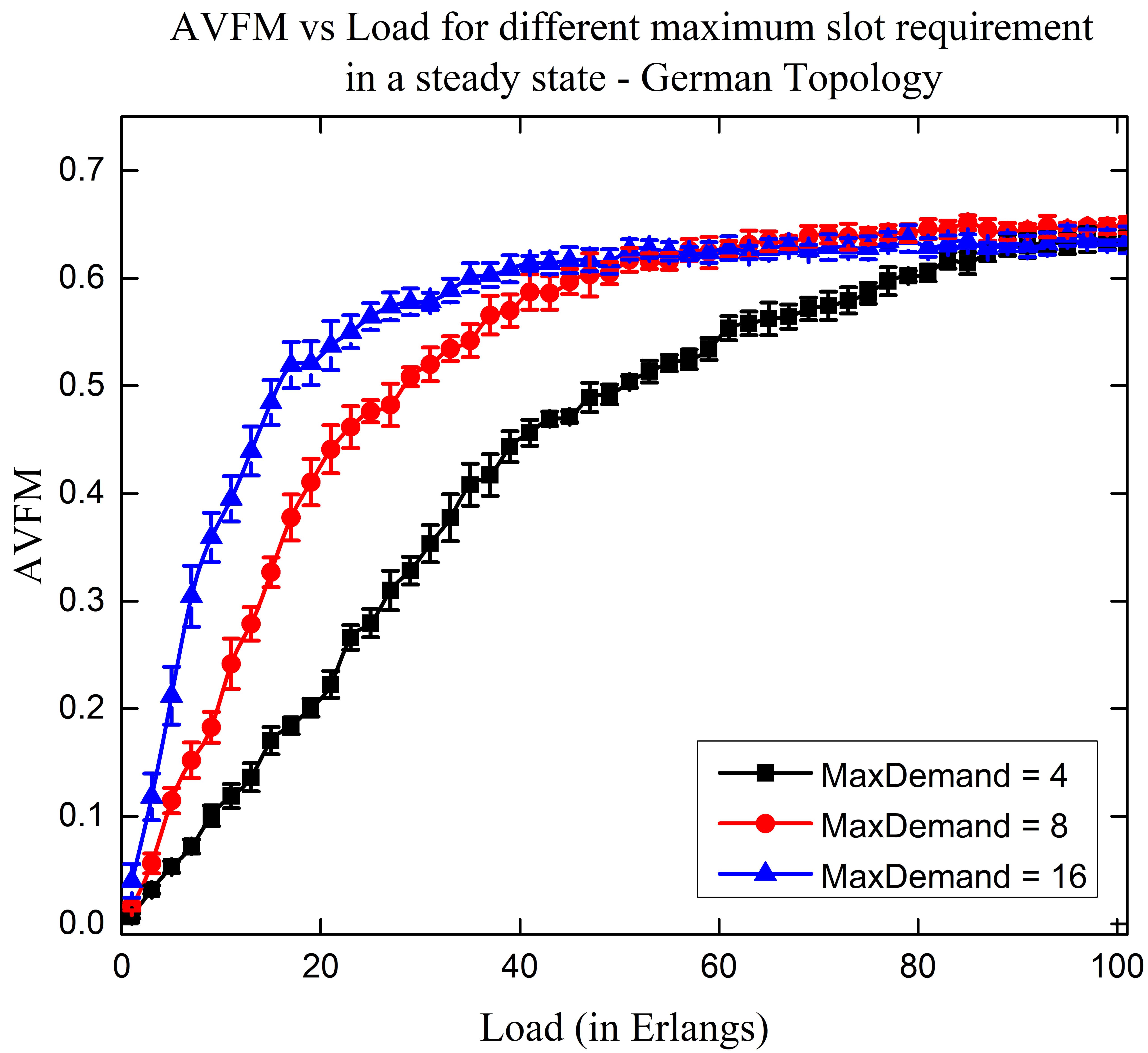}  
  \caption{}
  \label{fig:DMGER}
\end{subfigure}
\caption{AVFM vs traffic load in steady-state for (a) Net-A, (b) NSFNet, and (c) German topology for different maximum demand.}
\label{fig:fig12}
\end{figure*}

We also observe that the maximum permissible slot size or granularity range of the incoming connection requests (MaxDemand) also affect the fragmentation level (fig.\ref{fig:fig12}(a)-(c)). The higher granularity range (MaxDemand 8 and 16) indicates more fragmentation as the arriving connection requests are more diverse and leave the spectrum disorganized. We observe this behavior in all the three network topologies. However, at very high loads, the fragmentation level for such range may fall due to the unavailability of network slices fig.\ref{fig:fig12}(b)-(c).  

\begin{figure*}
\begin{subfigure}{.34\textwidth}
  \centering
  \includegraphics[width=.7\linewidth]{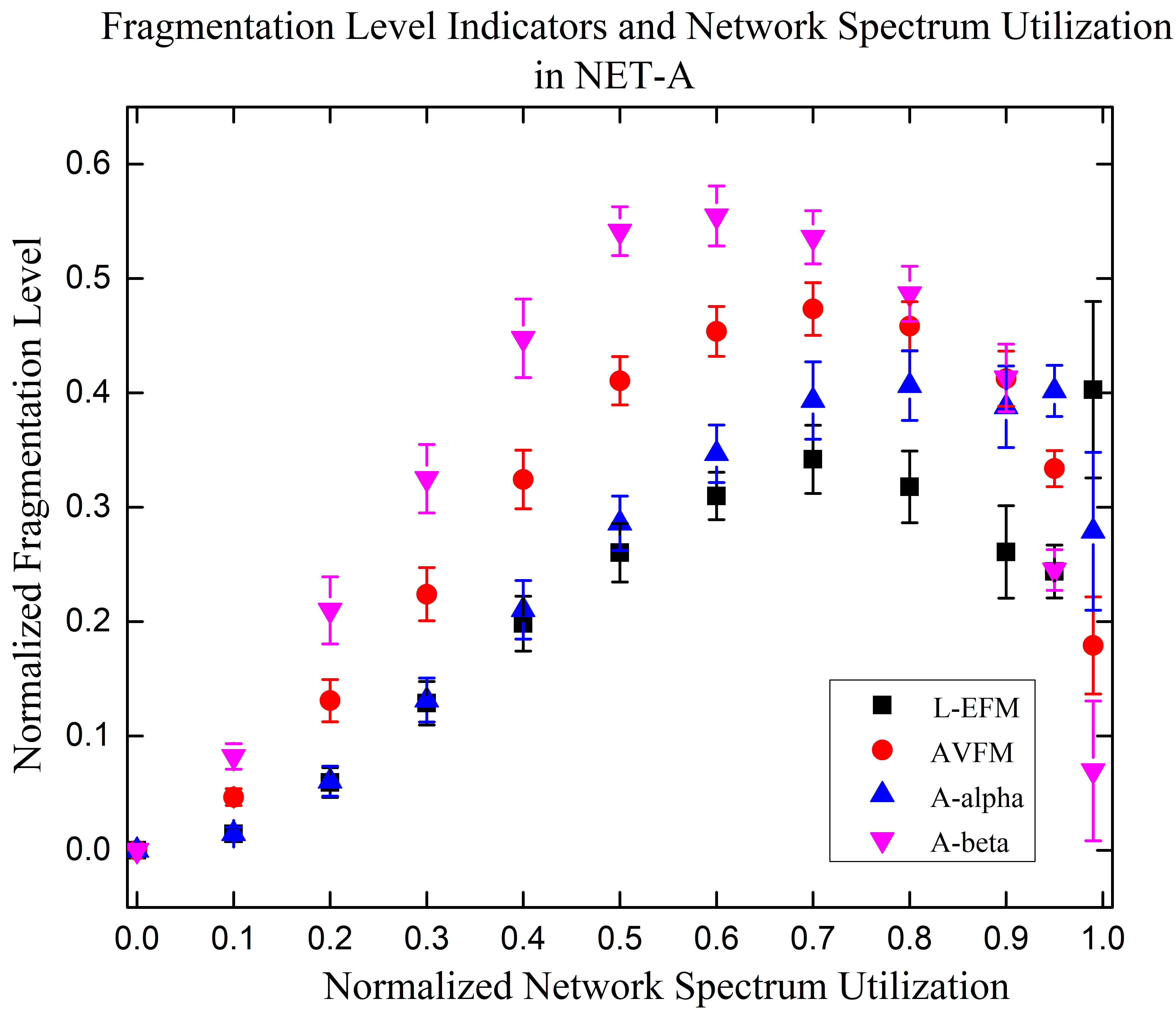}  
  \caption{}
  \label{fig:FigNETA_U}
\end{subfigure}
\begin{subfigure}{.34\textwidth}
  \centering
  \includegraphics[width=.7\linewidth]{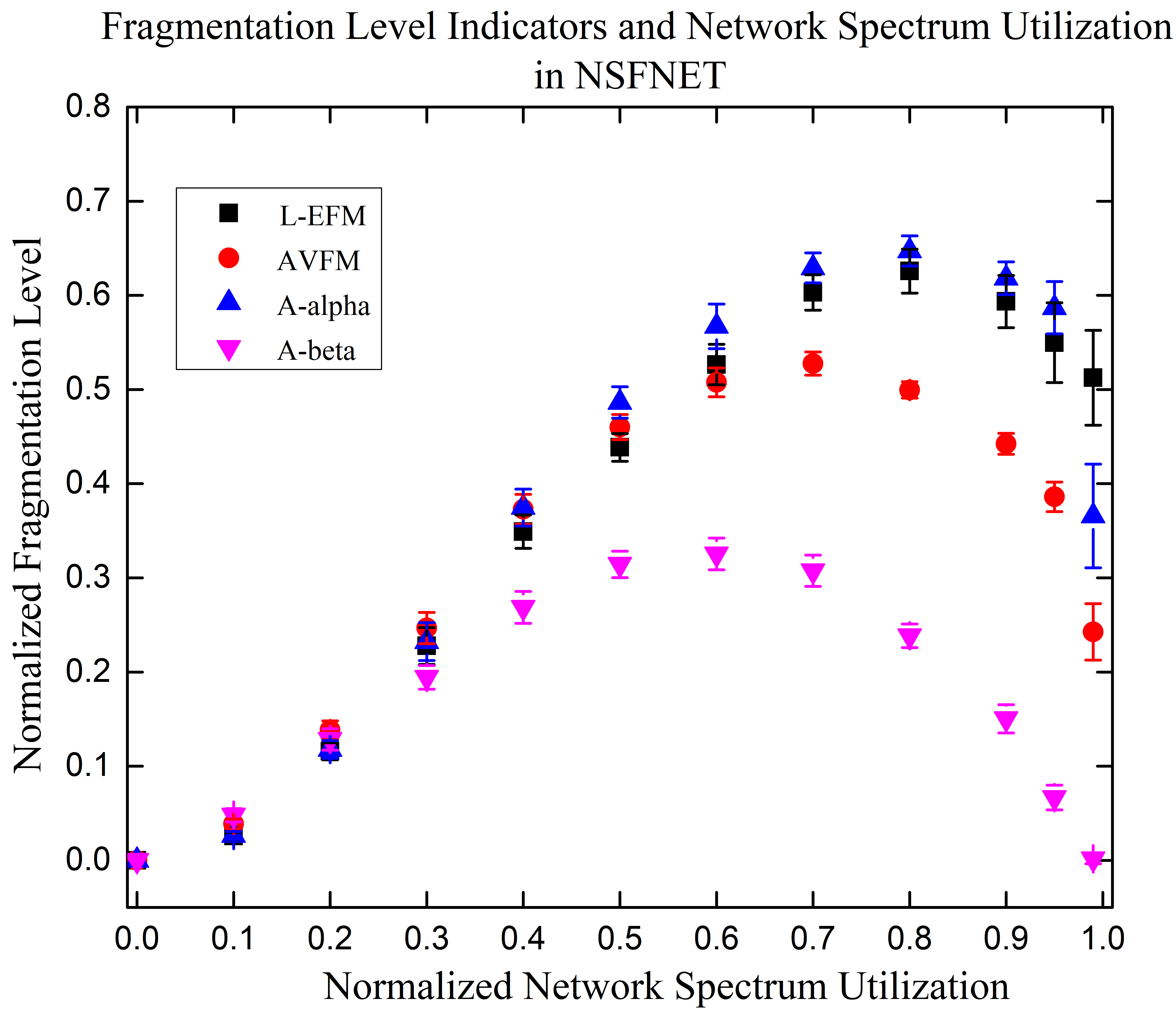}  
  \caption{}
  \label{fig:FigNSF_U}
\end{subfigure}
\begin{subfigure}{.34\textwidth}
  \centering
  \includegraphics[width=.7\linewidth]{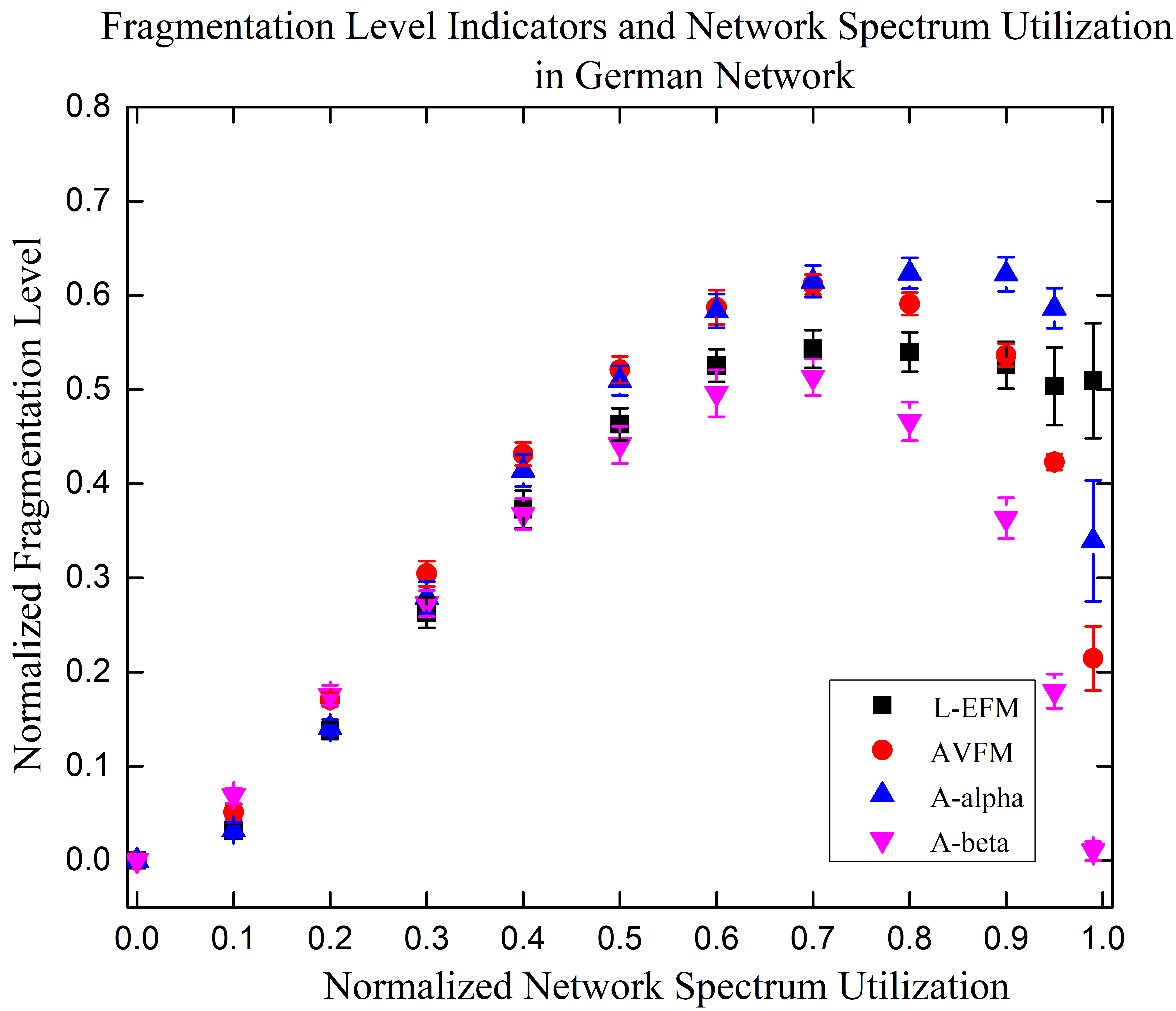}  
  \caption{}
  \label{fig:FigGER_U}
\end{subfigure}
\caption{Fragmentation level indicators with network utilization for (a) NET-A, (b) NSFNET and (c) German Network}
\label{fig:fig13}
\end{figure*}

In fig.~\ref{fig:fig13}, we also record the correlation between fragmentation level indicators and the normalized network spectrum utilization. We start with a completely vacant network spectrum (network utilization = 0.0) till the network spectrum is fully or almost fully occupied (network utilization  $\geq$0.99) for three network topologies. The L-EFM and A-alpha are positively correlated with network utilization. The AVFM and A-beta are moderately correlated with network utilization. We observe that the AVFM, the L-EFM, the A-alpha, and the A-beta are minimal for low network utilization, as expected. In fig.~\ref{fig:fig13}(a),(b), and (c)), at a low level of spectrum utilization, enough connection requests have not arrived, and sufficient contiguous and continuous spectrum slices are available for future requests. When the network spectrum utilization increases, so do the fragmentation level indicators increases. However, this condition does not remain same in high spectrum utilization cases. There are many active connection requests for the high spectrum utilization, and there are not enough available spectrum slices to cause significant fragmentation. Thus, fragmentation should be less in the high spectrum utilization case. AVFM seems to follow this condition, but L-EFM again ignores it due to non-consideration towards smaller fragment sizes. We can see in all network utilization plots ( fig.~\ref{fig:fig13}(a), (b), and (c)) that the A-beta captures the true essence of fragmentation in an empty and nearly full network spectrum. Hence, the contribution of the beta-component is significant in a true fragmentation indicator. The beta-component affects the network fragmentation level significantly; hence, it is better to consider the continuity fragmentation component in the overall analysis. 

 \section{Conclusion}
There has not been any metric using both continuity and contiguity aspect for network-level fragmentation in elastic optical network spectrum to the best of our knowledge. In this work, we used the vectored fragmentation metric to quantify the fragmentation level in a network at some instant. This work has assessed a vectored fragmentation metric's (AVFM) ability to capture continuity as well as contiguity fragmentation level in real-time network scenarios. We formulate the VFM to include  $1^{st}$-level continuity fragmentation metric for applications with a single slice requirement. If there is significant fragmentation for low-bandwidth (single slice) connection requests, it also represents a worst-case scenario for applications with large bandwidth requirements. We observed that fragmentation level evolves with time and network spectrum status. We observed the AVFM, the L-EFM, the A-alpha, and the A-beta over time. The continuity (A-beta) and contiguity (A-alpha) aspects contribute significantly to the overall fragmentation level. We also observe how the A-beta level is more at low traffic load, and the A-alpha surpasses it at some cross over point as load increases. The other network characteristics like arrival rate and the connection requests' granularity range also contribute to fragmentation metric (AVFM). We have also studied the correlation between the network spectrum utilization and fragmentation level. It is worth noting that AVFM captures the essence of fragmentation in empty as well as in fully utilized network spectrum more accurately than L-EFM.  

In future work, we plan to investigate the effectiveness of this metric in triggering the defragmentation procedure. We also plan to utilize the proposed vectored fragmentation metric in the joint Routing and Spectrum Provisioning (RSA). It can be used in optimizing routing decisions and slot selections to efficiently use the spectral resources. The work can be extended to include the number of paths in continuity assessment and the multiple slices based continuity fragmentation.  The individual components (multi-level fragmentation indicators) of the vectored fragmentation metric can be moulded to act as the application-specific indicators.

\end{document}